\PassOptionsToPackage{svgnames}{xcolor}
\documentclass[12pt]{article}

\usepackage{amsmath,
amssymb,
amsfonts,
amsthm,
setspace,
tikz,
 xcolor,
array,
}
\usetikzlibrary{calc}
\usetikzlibrary{positioning}
\usetikzlibrary{shapes.geometric}
\usetikzlibrary{patterns}
\usetikzlibrary{decorations.pathmorphing}
\usetikzlibrary{arrows.meta, positioning}
\usetikzlibrary{decorations.pathreplacing}
\usepackage[longnamesfirst]{natbib}
\usepackage{centernot}
\usepackage{xspace}
\usepackage{graphicx}
\usepackage{enumitem}
\usepackage[makeroom]{cancel}
\usepackage{bm}
\usepackage{bbding,pifont} 
\usepackage{graphicx} 
\usepackage{titlesec}
\usepackage{newunicodechar}

\usepackage[letterpaper]{geometry}
\definecolor{lam1}{HTML}{C70039}
\definecolor{lam2}{HTML}{2c7829}
\definecolor{lam3}{HTML}{3d2978}
\usepackage[colorlinks=true,urlcolor=lam2,linkcolor=lam2,citecolor=lam2]{hyperref}

\newtheoremstyle{mytheoremstyle}
    {\topsep} {\topsep} {\itshape} {} {\sc} {.} {.5em} {}
\theoremstyle{mytheoremstyle}
\newtheorem{definition}{Definition}

\newtheorem{theorem}{Theorem}
\newtheorem{lemma}{Lemma}
\newtheorem{proposition}[theorem]{Proposition}

\newcommand{\caxadornment}{}  

\newtheoremstyle{scfont}
    {\topsep} {\topsep} {} {} {\textbf} {} {.5em} {\textbf{\thmname{#1}\thmnumber{#2}}\thmnote{\ \ {\textsc{#3}.}}}
\theoremstyle{scfont}

\newtheorem{iax}{I}
\newtheorem{cax}{C}

\let\oldthecax\thecax
\renewcommand{\thecax}{\oldthecax\caxadornment}

\theoremstyle{remark}   
\newtheorem{remark}{Remark}
\newtheorem{example}{Example}

\newenvironment{tproof}[1]{%
    \phantomsection
    \label{pf:#1}%
    \addcontentsline{toc}{subsection}{Pf. Thm. \ref{#1}}%
    \begin{proof}[Proof of Theorem~\ref{#1}]%
}{%
    \end{proof}%
}
\newenvironment{pproof}[1]{%
    \phantomsection
    \label{pf:#1}%
    \addcontentsline{toc}{subsection}{Pf. Prop. \ref{#1}}%
    \begin{proof}[Proof of Proposition~\ref{#1}]%
}{%
    \end{proof}%
}

\newenvironment{subproof}[1][\proofname]{%
  \begin{proof}[#1]%
}{%
  \end{proof}%
}

\titleformat*{\section}{\large\bf\centering}

\titleformat{\subsection}[runin]
  {\bfseries}         
  {\thesubsection \quad }                  
  {0pt}               
  {}[.]               

\renewcommand{\emptyset}{\varnothing}
\renewcommand{\phi}{\varphi}

\def \<{\langle}
\def \>{\rangle}

\newunicodechar{￥}{\rmb}

\def \L{\mathcal{L}}
\def \LL{\mathcal{L}^\circ}
\def \LS{\mathcal{L}^{\ast}}
\def \LCom{\mathcal{L}^{COM}}
\def \LJnt{\mathcal{L}^{JNT}}

\def \I{\mathcal{I}}

\renewcommand{\implies}{\rightarrow}
\def \imp{\Rightarrow}
\def \V{\mathcal{V}}
\def \W{\Omega}
\def \w{\omega}

\def \T{\textup{T}}
\def \Tij{\textup{T}_{i \to j}}
\def \Tji{\textup{T}_{j \to i}}
\def \Tot{\textup{T}_{1 \to 2}}
\def \Tto{\textup{T}_{2 \to 1}}
\def \V{\textup{V}}
\def \Vij{\textup{V}_{i \to j}}
\def \Vji{\textup{V}_{j \to i}}
\def \Vot{\textup{V}_{1 \to 2}}
\def \Vto{\textup{V}_{2 \to 1}}
\def \l{\lambda}

\def \v{\mathbf{v}}
\def \ax{B}

\DeclareMathOperator*{\up}{\Uparrow^\star}

\makeatletter
\newcommand{\mylabel}[3]{\def\@currentlabel{#2}\phantomsection\textsc{#3} (\texttt{#2})\label{#1}}
\newcommand{\sclabel}[3]{\def\@currentlabel{\theax}\phantomsection\innercase{\textsc{#1}.}\label{ax:#2}}
\makeatother

\newcommand{\tref}[1]{\hyperref[#1]{\textcolor{lam1}{\textup{\textbf{\ref{#1}}}}}}
\newcommand{\cref}[1]{\hyperref[#1]{\textcolor{lam1}{\textup{\textbf{C\ref{#1}}}}}}
\newcommand{\iref}[1]{\hyperref[#1]{\textcolor{lam1}{\textup{\textbf{I\ref{#1}}}}}}
\newcommand{\rax}[1]{\hyperref[ax:#1]{\textup{(\textsc{#1})}}}

\newcommand{\gotopf}[1]{\noindent\hyperref[pf:#1]{\scalebox{1.3}{\raisebox{-0.3ex}{\ding{43}}}\,Go to Proof.}}

\title{\vspace{-8ex} Do You Know What I Mean?  \\ \large A Syntactic Representation for Differential
Bounded Awareness\thanks{%
This research was supported by an\ Australian Research Council Laureate
Felllowship. We wish to thank Edi Karni, Peter Wakker, Satoshi Fukuda, Emel
Filiz-Ozbay, as well as participants of the workshop on Unawareness in
Aarhus 2024, FUR 2024 in Brisbane and TUS X in Paris for helpful comments
and suggestions.}}
\author{Ani Guerdjikova\thanks{%
University of Grenoble-Alpes, CNRS, INRA, Grenoble INP, GAEL, 1241 rue des R%
\'{e}sidences, 38400 Saint Martin d'H\`{e}res, email: \texttt{%
ani.guerdjikova@univ-grenoble-alpes.fr}, Tel.: +33 4 56 52 85 78,} \and Evan
Piermont\thanks{%
Department of Economics, Royal Holloway, University of London, email: 
\texttt{evan.piermont@rhul.ac.uk}} \and John Quiggin\thanks{%
University of Queensland, School of Economics, Faculty of Business,
Economics and Law, email: \texttt{j.quiggin@uq.edu.au}, Tel.: +61 7 334 69646%
}}

\begin{document}

\maketitle

\begin{abstract}
\centering
\begin{minipage}{0.9\textwidth}  
\setstretch{.9}
Without the assumption of complete, shared awareness, it is necessary to consider communication between agents who may entertain different representations of the world. A syntactic (language-based) approach provides powerful tools to address this problem.  In this paper, we define translation operators between two languages which provide a ``best approximation'' for the meaning of propositions in the target language subject to its expressive power. We show that, in general, the translation operators preserve some, but not all, logical operations. We derive necessary and sufficient conditions for the existence of a joint state space and a joint language, in which the subjective state spaces of each agent, and their individual languages, may be embedded. This approach allows us to compare languages with respect to their expressiveness and thus, with respect to the properties of the associated state space.

\vspace{1cm}


\end{minipage}
\end{abstract}


\maketitle

\newpage

\section{Introduction}


Since the foundational work of \cite{savage1954foundations}, decisions under uncertainty have been modeled using a state-space, where each state represents a full resolution over uncertainty and where
preferences are defined over state-contingent acts, that is, mappings from the state space to a space of outcomes. 
However, the state-space approach has been subject to fundamental challenges from two directions. First, theories of bounded and differential awareness have challenged the assumption of a commonly known set of states of the world.\footnote{E.g., \cite{halpern2009reasoning,heifetz2006interactive,karni2013reverse,grant2013inductive,piermont2017introspective}.} In these theoretical frameworks, agents may be completely unaware of some possible states of the world (referred to as \emph{restricted} awareness) or may be unaware of relevant distinctions between states (\emph{coarse} awareness). 

Second, the direct use of a states to represent uncertainty has been criticized as being less natural than syntactic representations based on languages.\footnote{The classical representation of \cite{stone1936theory} establishes a correspondence between syntactic (language based) and semantic (state space) models, obviating the need to directly model syntax in the case of a single logically omniscient agent who aware of all possible events. Absent these assumptions, syntactic models may indeed capture novel considerations, see \cite{grant2013inductive,halpern2015syntax,blume2021constructive,piermont2026coarse,piermont2026failures}.}
When agents share the same language, they can also reason using the same state-space representation.
However, when agents' languages give rise to distinct perceptions of the world, an event as described by one agent may find no exact analog in another agent's language. As a result, the translation of a statement may yield \emph{ambiguity}, that is, a range of possible meanings in the target language.

In this paper, we directly model languages as the medium of communication
between agents (or between different selves of an agent, whose awareness may evolve over time).
We derive the normatively appealing properties of translation between languages; our main axiom states that the best possible translation yields the smallest range of meaning. That is, translation should not introduce more ambiguity than is strictly necessary. Our main results show that such a translation exists if and only if the distinct languages can be viewed as different (partial) descriptions of a universal space of uncertainty. Further, we show how such a translation might arise and how it can be used to delineate both common and distributed awareness of the speakers.

To illustrate how differential awareness leads to imprecise translation, consider two agents whose languages contain propositions about the price of oil, say barrels of Brent Crude, on a given day. Agent $1$ considers possible ranges for the price expressed relative to increments of 10 US dollars. For example, $\lambda^{70}$ represents the statement ``the price is strictly below \$70'' or $\lambda_{40}$ represents the statement ``the price is weakly above \$40'' (superscripts represent strict upper-bounds and subscripts weak lower bounds). By using compound statements, we can construct more complicated statements such as  $\lambda^{80}_{20} := (\lambda^{80} \texttt{ and } \lambda_{20})$: ``the price is dollars is in the interval $[20,80)$.''

Agent $2$, on the other hand, considers possible ranges for the price expressed in increments of 100 Chinese yuan (equivalent to exactly \$15 for the purposes of this example). So $\eta_{300}^{500}$ might be interpreted as ``the price is between \textyen300 and \textyen500 ". 
There are some propositions, such as $\lambda_{60}^{90}$ and $\eta_{400}^{600}$, that are exact translations of one another in the sense that from our external perspective, they refer to the same event. 
Of course, this is not always the case: the statement $\eta_{500}^{700}$ has no exact translation, there is no statement (compound or otherwise) in $1$'s language that can express exactly the same event. 


\subsection{Outline}

We consider two agents, $i \in \{1,2\}$, each with their own language $\L_i$. 
A language is a set of statements as well as the logical relationship between them; specifically we model the belief and awareness of agent $i$ though their subjective understanding of implication $\imp_i$ as given by $\L_i$. For two statements $\lambda, \lambda' \in \L_i$, say that $\lambda$ is \emph{more specific} than $\lambda'$ (equivalently, $\lambda'$ is \emph{more general} than $\lambda$) when $\lambda$ implies $\lambda'$.  For example, under the standard interpretation, $\lambda^{50}$ is more specific than $\lambda^{80}$, since in any state-of-affairs such that the former is true, the later is true as well.

We make no a priori assumption regarding the relation between the individual languages, and in particular, do not assume the existence of a universal state-space (or language) from which the agents' languages can be derived. 
And so, we must consider the problem of translation: how is one agent's description of an event (i.e., an event according to his own subjective understanding about how uncertainty might resolve) interpreted by another agent. Because some propositions in one language are not expressible in the other, translation may only \emph{approximate} the original meaning.

\begin{figure}[]
\centering
\begin{tikzpicture}[]

\def\h{1}
\def\sk{.2}

\node[] at (1,\h*.5) {$W_1$};
\node[] at (1,\h*.5 + \sk + \h) {$W_2$};

\draw[pattern=north east lines, pattern color=lam1!20, draw=none] (8.5,0) rectangle (12.5,\h);
\draw[pattern=north west lines, pattern color=lam2!30, draw=none] (9.5,0) rectangle (11.5,\h);

\draw [draw=black, fill=none, ultra thick] (1.5,0) rectangle (13.5,\h);
\foreach \i in {1,...,12}
{
\draw [draw=black, fill=none, thick] (\i+.5,0) rectangle (\i+1.5,\h);
  \pgfmathtruncatemacro{\dlabel}{(\i-1)*10}
  \pgfmathtruncatemacro{\ulabel}{\i*10}
\node at (\i+1,\h*.5) {$\lambda^{\ulabel}_{\dlabel}$};
}

\draw[draw=lam2, ultra thick] (9.5,0.05) rectangle (11.5,\h-.05);
\draw[draw=lam1, ultra thick] (8.5,0-.05) rectangle (12.5,\h+.05);

\draw[fill=lam3!15, draw=none] (9,\h+\sk) rectangle (12,\sk+\h*2);

\draw [draw=black, fill=none, ultra thick] (1.5,\h+\sk) rectangle (13.5,\sk+\h*2);
\foreach \i in {1,...,9}
{
\draw [draw=black, fill=none, thick] (\i*1.5,\h+\sk) rectangle (13.5,\sk+\h*2);
}

\foreach \i in {1,...,8}
{
  \pgfmathtruncatemacro{\dlabel}{(\i-1)*100}
  \pgfmathtruncatemacro{\ulabel}{\i*100}
\node at (\i*1.5+.75,\h*1.5 + \sk) {$\eta^{\ulabel}_{\dlabel}$};
}



\end{tikzpicture}
\caption{State-spaces for Agents $1$ and $2$ conversing about the price of Brent Crude. The event corresponding to $\eta_{500}^{700}$ = ``the price is between \textyen500 and \textyen700 " is highlighted in \textcolor{lam3}{blue}.
The inner and outer translations of $\eta_{500}^{700}$ are also demarcated, in \textcolor{lam2}{green} and \textcolor{lam1}{red}, respectively.}
\label{fig:oil}
\end{figure}

We address the problem of inexact translation by introducing the idea of inner and outer translation operators, functions from one language to the other. We provide three basic axioms of translation which yield the following interpretation of the translation operators:
When translating a statement $\eta \in \L_j$ from a source language to a target language, the inner translation of $\eta$ will be the most general statement in the target language more specific than $\eta$.\footnote{The outer and inner translation have their equivalents in linguistics. A \emph{hyponym} is a word which represents a special case of another word: `dog' is a hyponym of `animal'. A \emph{hypernym} is a word which subsumes the meaning of another word: `animal' is a hypernym of `dog'. The inner translation is most general hyponym of $\eta$ in the target language; the outer translation,  the most specific hypernym.}  Conversely, the outer translation will be the most specific proposition in the target language more general than $\eta$. 

In our example, the inner translation of $\eta_{500}^{700}$ is the statement $\lambda_{80}^{100}$, since whenever the later is true, so is the former (it is more specific), and $i$'s language has no way to  generalize $\lambda_{80}^{100}$ without allowing for prices that are precluded by $\eta_{500}^{700}$. In dual fashion, the outer translation is the statement $\lambda_{70}^{110}$, since whenever the former is true, so is the later (it is more general), and $i$'s language has no way to  specialize $\lambda_{70}^{110}$ without precluding prices that are allowed by $\eta_{500}^{700}$. The inner and outer translation are shown in Figure \ref{fig:oil}.


So far,  the linguistic constraints of the two agents simply apportion the world into distinct partitions of events, but otherwise refer to the same possibilities. There is, however, another sort of incongruity that can arise due to differential awareness: one agent may be unaware of some resolution of uncertainty, in the sense that none of his statements allow for its possibility.  For example, suppose that agent $2$, but not $1$, is aware of the possibility that oil prices can be negative, as happened briefly in the early stages of the Covid pandemic. Where $1$ sees $\lambda^0$ as a contradiction (it is impossible for the price to be strictly less than \$0), agent $2$ does not entertain an analogous belief. So, while $\eta^{200}$ allows, from agent $2$'s perspective, for the possibility of negative prices, there is no way to maintain this allowance after translation.  As such, the inner translation of $\eta^{200}$ will be $\lambda^{30}$, and, more interestingly, the outer translation is undefined, since there does not exist \emph{any} statement in agent $1$'s language which would generalize $\eta^{200}$ insofar as it would  allow negative prices.\footnote{Formally, we define the outer translation as $\ast$, some all encompassing state-of-affairs that includes potential unknowns lying outside of $i$'s language.}

\begin{figure}[]
\centering
\begin{tikzpicture}[]
\def\h{1.2}
\def\sk{.2}
\node[] at (-.6,\h*.5) {$W_1$};
\node[] at (-.6,\h*.5 + \sk + \h) {$W_2$};

\draw[pattern=north west lines, pattern color=lam2!30, draw=none] (1.5,0) rectangle (4.5,\h);
\draw [draw=black, fill=none, ultra thick] (1.5,0) rectangle (13.5,\h);
\foreach \i in {1,...,12}
{
  \draw [draw=black, fill=none, thick] (\i+.5,0) rectangle (\i+1.5,\h);
  \pgfmathtruncatemacro{\dlabel}{(\i-1)*10}
  \pgfmathtruncatemacro{\ulabel}{\i*10}
  \node at (\i+1,\h*.5) {$\lambda^{\ulabel}_{\dlabel}$};
}
\draw[draw=lam2, ultra thick] (1.5,0+.05) rectangle (4.5,\h-.05);

\draw[decorate, thick, lam1, decoration={brace, amplitude=6pt, mirror}]
  (0,-.2) -- (13.5,-.2)
  node[midway, below=6pt] {\Large $\ast$};

\draw[fill=lam3!15, draw=none] (0,\h+\sk) rectangle (4.5,\sk+\h*2);
\draw [draw=black, fill=none, ultra thick] (0,\h+\sk) rectangle (13.5,\sk+\h*2);
\foreach \i in {0,...,9}
{
  \draw [draw=black, fill=none, thick] (\i*1.5,\h+\sk) rectangle (13.5,\sk+\h*2);
}
\foreach \i in {1,...,8}
{
  \pgfmathtruncatemacro{\dlabel}{(\i-1)*100}
  \pgfmathtruncatemacro{\ulabel}{\i*100}
  \node at (\i*1.5+.75,\h*1.5 + \sk) {$\eta^{\ulabel}_{\dlabel}$};
}
\node at (.75,\h*1.5 + \sk) {$\eta^{0}$};
\end{tikzpicture}
\caption{The event corresponding to $\eta^{200}$ = ``the price is between strictly less than \textyen200" is highlighted in \textcolor{lam3}{blue}. The inner translations of $\eta^{200}$ is demarcated in \textcolor{lam2}{green}; the outer translation is defined as $\ast$ some all encompassing state-of-affairs that includes potential unknowns, demarcated in \textcolor{lam1}{red}.}
\label{fig:oil-2}
\end{figure}

We then ask the following central question.
When can the agents' individual perceptions of uncertainty be viewed as bounded awareness of a \emph{joint state space}? That is, when can the statements in both of the agents' languages be simultaneously (and consistently) interpreted as describing events in a single state-space?
Our main result shows that the existence of such a joint semantics is exactly characterized by existence of a translation satisfying our axioms. This serves as both a normative justification for our definition of translation as well as an explicit construction of the joint state-space, given a consistent translation.
We use this result to characterize comparative awareness based on the translation operators, even when the languages used by two agents are non-nested.

We next consider the possibility of exact translation, arising when the inner and outer translations coincide. The set of propositions for which translation is exact form a lattice which we describe as the common language. In the special case when one language is strictly more expressive than another, we show that the less expressive language is isomorphic to the common language, while the more expressive language is is.omorphic to the joint language

When considered from the semantic perspective of a joint state-space, the events expressed by the inner and outer translations correspond to events contained in, and containing, the event referred to by the original proposition; specifically, these are lower and upper approximations in the sense of \cite{dempster1967upper}. If the agents entertain a probability over the state-space, then the probability of that $\lambda$ being true is equal to the probability of the event it describes---so, when $\lambda$ is translated, the speaker of the target language will entertain a range of probabilities for $\lambda$: the lower-bound probability being the probability corresponding to the inner translation of $\lambda$ (i.e., lower approximation) and the upper-bound that of the outer translation (i.e., upper approximation). Hence, translations serves as a bridge between differential awareness and ambiguity (now in the decision theoretic sense).

\vspace{-2ex}

$$
 \blacklozenge \  \blacklozenge \  \blacklozenge 
$$

\vspace{0ex}

The next section introduces the primitive elements of our model, namely, the formalities of languages and state spaces. Section \ref{sec:translation_between_languages} introduces our framework of translation and our three normative criteria for consistent translation. Section \ref{sec:implication} then shows that consistent translation can be fully characterized by a cross-language implication operator. Our main representation results are in Section \ref{sec:joint-ss}, which shows that consistent translation is equivalent to the existence of a joint state-space. We describe the common language in Section \ref{sec:common-ss}. We conclude with Section \ref{sec:lit}, containing a discussion of related literature.

\section{Preliminaries }

\subsection{Languages}

A \emph{language} $\L$ is a pair: $\L = (L,\ax)$, where $L$ is a set, $B$ is a subset of $L$, and $L$ is closed under negation $\neg$, disjunction $\lor$, and conjunction $\land$. That is, if $l$ and $l'$ are two elements of $L$, then so too are $\neg l$, $l \lor l'$, and $l \land l'$. $B$ is a set of statements \emph{believed to be true}. In the example above, both agents believed that the price may not simultaneously take two distinct values and thus, statements such as $\neg ((\lambda^{100}_{90})\wedge(\lambda^{80}_{70}))$ or $\neg((\eta^{100}_{0}) \land (\eta^{300}_{200}))$ are believed to be true, and so, would be contained in the respective sets of beliefs $B_1$ and $B_2$. Agent 1 also believed that negative prices were impossible, and thus $\lambda_0 \in B_1$.

 A \emph{truth assignment} for $\L$ determines the truth for each statement in the languages; formally, a truth assignment is a function
$
w: L \rightarrow \left \{ 0,1\right \}
$
such that for all $l, l' \in L$:
\begin{enumerate}[label=(\roman*)]
    \item $w(\neg l) = 1 - w(l)$, $w(l \lor l') = \max\{w(l),w(l')\}$, and $w(l \land l') = \min\{w(l),w(l')\}$,
    \item $w(l) = 1$ for all $l \in \ax$.
\end{enumerate}

The interpretation is that $0$ represents false and $1$ true. As such, (i) requires that the assignment preserves truth with respect to the logical operators in the language and (ii) that it is consistent with the beliefs, insofar as the propositions $\ax \subseteq L$ must be assigned to be true.
Let $W(\L)$ collect the set of all truth-assignments for $\L$.

The language $\L$ is \emph{free of logical contradiction} if $W\left( \L \right) \neq \emptyset$. Each $w \in W\left( \L \right)$ is called a \textit{model} for $\L$.  In the semantic interpretation of the language, a model corresponds to a (subjective, to the speaker of the language) state and $W(\L)$ the state-space.\footnote{A special case is where  $L$ is built up from a finite set $L^{e}$ of elementary propositions, in which case we say it is finitely generated. When $L$ is finitely generated, then specifying a truth assignment $w$ over the set $L^e$ uniquely determines its value over all of $L$.}

Say that $l$ \textit{implies} $l^{\prime }$ in $\L$ if for all $w \in W\left( \L \right)$, $w(l) \leq w(l')$; in this case we write $l \imp_{\L} l^{\prime }$. If $l \imp_\L l'$, then whenever $l$ permits some state-of-affairs, $l'$ too permits it; for example, $l = $ `The shape is a square' and $l' = $ `The shape is a rectangle'.  As such, we say that $l$ is \emph{more specific} than $l'$ and that $l'$ is \emph{more general} than $l$. It is immediate that $\imp_\L$ is a partial order. 

 A proposition $l$ is a \textit{tautology} (receptively, \emph{contradiction}) for $\L$ if $w\left( l\right) =1$ (resp. $w\left( l\right) =0$) for all $w\in W\left( \L \right)$; the tautologies and contradictions are the most general and most specific statements, respectively, of the partial order $\imp_\L$. Call $\L$ \emph{complete} if every subset of $\L$ has a supremum and infimum (according to $\imp_\L$). 

Further, two propositions $l$ and $l^{\prime }$ are \textit{logically equivalent} in $\L$ if, for all $w \in W\left( \L \right)$, $w(l) = w(l^{\prime })$, in which case we write $l\Leftrightarrow_{\L} l^{\prime }$.
The operator $\Leftrightarrow_{\mathcal L}$ defines an equivalence relation on $\L$; let $\LL$ denote resulting quotient space, 
$\LL=\raisebox{.1em}{$L$}/ \raisebox{-.1em}{$\Leftrightarrow_{\L}$}$, called the Lindenbaum-Tarski algebra of $\L$.

It is well known that $\LL$ is a Boolean algebra that inherits its algebraic
structure from $\L$. As such, we extend our
logical operators (negation, disjunction, conjunction) implication to $\LL$: for $\lambda, \lambda' \in \LL$, $\lambda \imp_\L \lambda'$ iff $l \imp_\L l'$ for any (hence all) $\l \in \lambda$ and $%
l' \in \lambda'$. Similarly, truth valuations are extended to $\LL$ by setting $w(\lambda)
=w(l) $ for $l  \in \lambda$.\footnote{%
Taking the Boolean algebra $\LL$ as a primitive object, we can define $W(\L) $ to be the set of Boolean homomorphisms into the two element
Boolean algebra $\left \{ 0,1\right \} $. It is easy to show that this set
coincides with the set of models of any language with quotient
algebra $\LL$.} Not only is the quotient of a language $\L$ a Boolean algebra,
but in fact, every Boolean algebra arises this way:

\begin{remark}
\label{lemma-algebra-to-lang} Let $BA$ be a Boolean algebra. Then
there exists some language $\L$
such that $BA$ is isomorphic to $\LL$.\footnote{See for example, Theorem 14.4 of \cite{sikorski1969boolean}.}
\end{remark}

Remark \ref{lemma-algebra-to-lang} implies that using the set of equivalence
classes $\LL$ as a primitive is equivalent to using a languages $\L = (L,B)$ (under the maintained assumption of
logical omniscience). 

Because we want to allow for the possibility that a translation from one language to another is undefined, we require one additional ingredient. For a Lindenbaum-Tarski algebra $\LL$, define $\LS$ to be a lattice with elements $\LL \cup \{\ast\}$. 
Over $\LL$, the lattice ordering coincides with $\imp_{\L}$, which we extend to $\ast$ (in a slight abuse, without changing  notation) via 
 $$
 \lambda \imp_{\L} \ast \qquad \text{and} \qquad  \ast \  {\centernot{\imp}}_{\L}  \ \lambda \qquad \qquad \text{for all} \qquad \lambda \in \LL.
 $$ 
 In particular, this forces $\lambda \lor \ast = \ast$ and $\lambda \land \ast = \lambda$ for all $\lambda \in \LL$. The interpretation is that $\ast$ represents an `all-encompassing' event that includes everything the speaker of $\L$ is aware of and possibly more---of course, there is no way to subdivide this event beyond what can already be expressed by $\L$, and in particular, it may have no negation.

In the sequel, we consider two agents and so take as the central objects of our study two languages: $\L_i = (L_i,\ax_i)$ for $i \in \{1,2\}$.
In a slight abuse of notation, for each langues $\L_i$, let $\imp_i$ denote the associated implication operator, $\LL_i$ the corresponding Lindenbaum-Tarski algebra, $\LS_i$ its $\ast$-augmentation, and $t_i$, $f_i$, and $\ast_i$ the class of tautologies, the class of tautologies contradictions, and the 1-point extension, respectively.  In light of Remark \ref{lemma-algebra-to-lang}, we often work directly with these algebraic objects, under the understanding that they arise from some language $(L_i,\ax_i)$.

\section{Translation Between Languages}
\label{sec:translation_between_languages}

 In this section, we consider the problem of translation, positing general properties that any ideal translation must satisfy.  It will turn out (see Theorem \ref{thm:state-space-existance}) that these conditions amount exactly to the existence of a joint state-space for which each individual language serves as an incomplete description. 
Loosely speaking, we will define a translation operator as a function associating each statement in one language with a statement in the other. Of course, because equivalent statements ought to be treated equivalently, we operationalize this as a function from  
 $\LS_i$ to $\LS_j$. 

Formally, a \emph{translation operator from $i$ to $j$} is a function $\Tij: \LS_i \to \LS_j$. For $\lambda \in \LL_i$, $\Tij(\lambda)$ represents the translation of (the propositions in the equivalence class) $\lambda$ into the language $j$. For a translation operator $\Tij$ we refer to $i$ as the source language and $j$ as the target language. The principal element of our theory will be a \emph{translation} 
$$\T = \<\Tot^{-},\Tot^{+},\Tto^{-},\Tto^{+}\>$$
 consisting of four translation operators, two moving from $1$'s language into $2$'s and two in the opposite direction.

The operators $\Tot^{-}$ and $\Tto^{-}$ are referred to as \emph{inner} translation operators, and are intended to capture the most general statement in the target language that is still more specific than the statement being translated; in dual fashion,  $\Tot^{+}$ and $\Tto^{+}$ are \emph{outer} translation operators, capturing the most specific statement in the target language more general than the statement being translated. For example, if $\lambda \in \LL_i$, then $j$'s inner approximation of $\lambda$, given by $\Tij^{-}(\lambda)$, is the most general statement in $\LL_j$ that refers to a specification of whatever $\lambda$ refers to. Likewise, $j$'s outer approximation of $\lambda$, given by $\Tij^{+}(\lambda)$, is the most specific statement in $\LL_j$ that refers to a generalization of whatever $\lambda$ refers to.\footnote{We interpret $\Tij(\lambda_i) = \ast_j$ to indicate that the translation of $\lambda_i \in \LL_i$ is undefined. The idea here being that when the $j$ is unaware of something $\lambda_i$ refers to, then the outer translation of the $\lambda_i$ into $j$'s language will be undefined since there is no statement more general than $\lambda_i $.} 

We now introduce three conditions on a translation that capture this interpretation outlined above. These conditions should be understood to hold for any $i,j \in \{1,2\}$ with $i \neq j$:
 
\begin{cax}[Galois]
\label{c:galois}
For all $\lambda_i \in \LS_i$ and $\eta_j \in \LS_j$:  
\begin{equation}
\label{eq:galois}
\tag{$\mathfrak{g}$}
\eta_j \imp_j \Tij^{-}(\lambda_i) \qquad \qquad \text{ if and only if}  \qquad \qquad \Tji^{+}(\eta_j) \imp_i \lambda_i.
\end{equation}
\end{cax}

\cref{c:galois} states the the inner and outer translations moving in opposite directions are Galois connections of one another. This is a principle of optimality, requiring that the outer translation is maximally efficient given the inner one, and vice versa. To get a sense of this, consider the translations in Figure \ref{fig:violation}: $\lambda$ is revealed to be more general than $\eta$, since its inner translation lies above $\eta$. $\Tji^{+}(\eta)$ is suppose to be the most specific statement in $i$'s language more general than $\eta$, but since $\Tji^{+}(\eta) \centernot{\imp}_i \lambda$ it follows that $\Tji^{+}(\eta) \land \lambda$ (show in \textcolor{lam2}{green}) is strictly more specific than $\Tji^{+}(\eta)$ and, as both constituent parts are, it is also more general than $\eta$.

\begin{figure}[]
\centering
\begin{tikzpicture}[
  scale=1.0,
  >=Stealth,
  edge/.style={-{>[scale=2.3,length=1.4,width=1.4]}, line width=1pt},
  trans/.style={-{>[scale=2.0,length=3.2,width=3.2]}, line width=1.5pt, dashed},
  node/.style={
    draw,
    rounded corners=3pt,
    minimum height=7mm,
    minimum width=12mm,
    fill=black!5,
    align=center,
    font=\small,
    line width=1pt
  },
  dots/.style={font=\Large}
]

\begin{scope}[shift={(-3.2,0)}]

  \node[dots] (ldown) at (0,0) {$\vdots$};

  \node[node] (LA)  at (0,1.2) {$\eta$};
  \node[node] (LB)  at (-1.2,2.7) {}; 
  \node[node] (LBp) at ( 1.2,2.7) {$\Tij^{-}(\lambda)$};

  \node[node] (LC)  at (0,4.2) {$\Tij^{+}(\lambda)$};

  \node[dots] (lup) at (0,5.4) {$\vdots$};

  \draw[edge] (LA) -- (LB);
  \draw[edge] (LA) -- (LBp);

  \draw[edge] (LB) -- (LC);
  \draw[edge] (LBp) -- (LC);

  \draw[edge] (ldown) -- (LA);
  \draw[edge] (LC) -- (lup);

\end{scope}

\begin{scope}[shift={(3.2,0)}]

  \node[dots] (rdown) at (0,0) {$\vdots$};

  \node[node, ultra thick, draw=lam2, fill=lam2!20] (RA)  at (0,1.2) {$\Tji^{-}(\eta)$};
  \node[node] (RB)  at (-1.2,2.7) {$\lambda$};
  \node[node] (RBp) at ( 1.2,2.7) {$\Tji^{+}(\eta)$};

  \node[node] (RC)  at (0,4.2) {};

  \node[dots] (rup) at (0,5.4) {$\vdots$};

  \draw[edge] (RA) -- (RB);
  \draw[edge] (RA) -- (RBp);

  \draw[edge] (RB) -- (RC);
  \draw[edge] (RBp) -- (RC);

  \draw[edge] (rdown) -- (RA);
  \draw[edge] (RC) -- (rup);

\end{scope}


\draw[trans, color=lam1] (LA) -- (RA);
\draw[trans, color=lam1] (LA) -- (RBp);

\draw[trans,color=lam1] (RB) -- (LBp);
\draw[trans,color=lam1] (RB) -- (LC);

\end{tikzpicture}
\caption{A violation of \cref{c:galois}, with translations as dotted lines.}
\label{fig:violation}
\end{figure}

Our next condition, \cref{c:approximation}, is more straightforward; it requires that the inner translation is indeed `inner' and the outer `outer' as given by the specificity ordering $\imp_i$.

\begin{cax}[Approximation]
\label{c:approximation}
For all $\lambda_i \in \LS_i$ we have $\Tij^{-}(\lambda_i) \imp_j \Tij^{+}(\lambda_i)$.
\end{cax}

Finally, our third condition \cref{c:r_dual} states that the translation preserves negation as ably as possible given the agents' differing awareness.

\begin{cax}[Restricted Duality]
\label{c:r_dual}
For $\lambda_i \in \LL_i$ such that $\Tij^{+}(\lambda_i) \neq \ast_j$
$$
\Tij^{-}(\neg \lambda_i) = \neg \Tij^{+}(\lambda_i) \land \Tij^{-}(t_i).
$$
\end{cax}

If the two agents agree on the space of possibilities (i.e., if $\Tij^{-}(t_i) = \Tij^{+}(t_i) = t_j$ and vice versa), then \cref{c:r_dual} simply requires that negation commutes with the translation. However, when $j$ is unaware of something $i$ is aware of, then the negation of $\lambda$ in $i$'s language will include everything $i$ is aware of, whereas the negation of its translation into $j$'s language will include only what $j$ is aware of. As such, the duality between the operators must be modulated by the most general statement $j$ is aware of: $\Tij^{-}(t_i)$.

Call a translation \emph{consistent} if \cref{c:galois}, \cref{c:approximation}, and \cref{c:r_dual} hold.

\subsection{Properties of Consistent Translation}

Despite its abstract nature, \cref{c:galois} provides a substantial structure to the way translation works. 
We here now consider some derived properties of translation operators.
In particular, say that $\Tij$

\begin{enumerate}[label=\textbf{T\arabic*}., ref=\textbf{T\arabic*}]
  \item \label{t:contradiction} 
  is \textit{extreme} if $\Tij(f_{i})  = f_{j}$ and $\Tij(\ast_i)  = \ast_j$,

  \item \label{t:concreteness} 
  is \textit{concrete} if $\Tij(\lambda_i) = \ast_j$ implies $\lambda_i = \ast_i$,

  \item \label{t:monotonicity} 
  \textit{preserves implication} if $\lambda \imp_i \lambda'$ implies $\Tij(\lambda) \imp_j \Tij(\lambda')$,

  \item \label{t:conjunction} 
  \textit{preserves conjunction} if $\Tij(\lambda \land \lambda') = \Tij(\lambda) \land \Tij(\lambda')$,

  \item \label{t:disjunction} 
  \textit{preserves disjunction} if $\Tij(\lambda \lor \lambda') = \Tij(\lambda) \lor \Tij(\lambda')$,
\end{enumerate}
for all $\lambda, \lambda' \in \LS_i$.

\begin{proposition}
\label{prop:equivalence}
Let $\T$ satisfy \cref{c:galois}. Then 

\begin{enumerate}[label=(\roman*)]
  \item the inner translation operators, $\Tot^{-}$ and $\Tto^{-}$, satisfy \tref{t:contradiction}--\tref{t:conjunction}.
  \item the outer translation operators, $\Tot^{+}$ and $\Tto^{+}$, satisfy \tref{t:contradiction}, \tref{t:monotonicity}, and \tref{t:disjunction}.
  \item the set of $\I(\Tij) = \{\lambda \in \LL_i \mid \Tij(\lambda) \neq \ast_j \}$ of `approximable' statements, is an ideal (i.e., directed downset).
\end{enumerate}
\gotopf{prop:equivalence}
\end{proposition}


\section{Cross Language Implication}
\label{sec:implication}

We motivated the inner and outer translations as being approximations of meaning built from those terms in the target language that are specific, and more general, respectively, than the translated statement. For example, the inner translation captures the most general statement in the target language that is still more specific than the statement being translated. Of course, this raises the question as to when a statement in one language is more specific than another in a different language? Indeed, this interpretation of our translation operators requires that there is an implication structure across languages, revealing some underlying relation between the references. Thus, to better understand the structure of translation as imposed by our consistency axioms, we directly examine the notion of \emph{cross language implication}. In this section, we show that such a cross language implication operator completely characterizes the theory of consistent translation laid out in Section \ref{sec:translation_between_languages}. 

Formally, define $\imp^{\star}$, a binary relation over $\LS_1 \cup \LS_2$, that describes implication \emph{across} languages.
Below, we provide five properties of $\imp^\star$ that capture consistent translation: the first two are coherency postulates which provide the structure needed to interpret cross language implication as a representation of the shared meaning in the two languages, extending the partial view held by each of the two agents.  The final three properties require that cross language implication preserves the meaning of the structural elements of the languages (i.e., $\lor$, $\land$, etc.). Consider the following properties of $\imp^\star$ where $i,j \in \{1,2\}$ and $i \neq j$:

\begin{iax}[Extensibility]
\label{i:extensibility}
Let $\lambda_i, \lambda'_i \in \L_i$:
$$\lambda_i \imp^{\star} \lambda'_i \qquad \text{ if and only if  } \qquad \lambda_i \imp_i \lambda'_i.$$
\end{iax}

\iref{i:extensibility} is a fundamental notion of coherence. By requiring that $\imp^\star$ extends the each $\imp_i$, \iref{i:extensibility} justifies our interpretation of cross language implication as reflecting the individual agents' understanding of their own languages.

\begin{iax}[Transitivity]
\label{i:transitivity}
$\imp^{\star}$ is transitive
\end{iax}

\iref{i:transitivity} is a basic tenet of logical structure---essentially a dictate of deductive closure---requiring internal consistency of the cross language implication. Indeed, if some shared meaning did exist, then the transitivity of implication is required to entertain any reasonable semantic interpretation.

The next three properties ensure that the cross language implication respects the structural properties of the language, the bounds, conjunction and disjunction, and negation, respectively: 

\begin{iax}[Bound Consistency]
\label{i:explosion}
$f_{i} \imp^\star f_j$ and $*_i \imp^\star *_j$
\end{iax}

Notice that, \iref{i:explosion} allows for the possibility that $t_{i} \ {\centernot{\imp}}^\star\  t_j$, and indeed, this is the case when $j$ is aware of some referent that $i$ is not. 

\begin{iax}[Connective Consistency]
\label{i:connective}
Let $H_j \subseteq \LL_j$ and $\lambda_i \in \LL_j$. Then:
\begin{enumerate}[label=(\roman*)]
  \item if $\bigwedge H_j$ exists, then $\lambda_i \imp^{\star} \eta_j$ for all $\eta_j \in H_j$ implies $\lambda_i \imp^{\star} \bigwedge H_j$
  \item if $\bigvee H_j$ exists, then $\eta_j \imp^{\star} \lambda_i$ for all $\eta_j \in H_j$ implies $\bigvee H_j \imp^{\star} \lambda_i$ 
\end{enumerate}
\end{iax}

\iref{i:connective} requires that $\imp^\star$ preserves the meets and joins of the lattices $\LS_i$ and $\LS_j$, and as such, the meaning of $\land$ and $\lor$ is preserved. Finally, \iref{i:neg_cons} requires also that the meaning of $\neg$ is preserved, albeit with the caveat that negation can be distorted by differential awareness:

\begin{iax}[Negation Consistency]
\label{i:neg_cons}
For $\lambda_i \in \LL_i$ and $\eta_j \in \LL_j$ such that $\lambda_i \imp^\star t_j$
$$
\eta_j \imp^\star \neg \lambda_i \qquad \text{ implies } \qquad \lambda_i \imp^\star \neg \eta_j.
$$
\end{iax}

When $t_i \imp^\star t_j$ and $t_j \imp^\star t_i$, then \iref{i:neg_cons} exactly states that $\imp^\star$ respects negation in the expected, i.e., classical, way. However, when agents are aware of different possibilities (i.e., when $t_{i} \ {\centernot{\imp}}^\star\  t_j$), then negation operates differently in different languages as it can `pull in' extra possibilities on agent was unaware of. Consider the following example:

\begin{example}
 Consider an English speaker $1$ and a Spanish  speaker $2$ discussing animals, all of whom are either mammals (\emph{mam\'ifero}) or who lay eggs (\emph{pone huevos}). For $1$, these events are described by $\lambda_{mam}$ and $\lambda_{egg}$, respectively, and for $2$, by $\eta_{mam}$ and $\eta_{huev}$. Both speakers agree that the animals under inquiry are one or other other, so that $(\lambda_{mam} \lor \lambda_{egg}) \Leftrightarrow_1 t_1$ and $(\eta_{mam} \lor \eta_{huev}) \Leftrightarrow_2   t_2$.
Further, assume that $1$, but not $2$, is aware of platypus ($\lambda_{plat}$), so that $(\lambda_{mam} \lor \lambda_{egg}) \Leftrightarrow_1 \lambda_{plat} \ {\centernot{\imp}}_{1} \ f_1$ but $(\eta_{mam} \land \eta_{huev}) \Leftrightarrow_2  f_2$.

The agents' languages are represented on the right and left sides of Figure \ref{fig:cross-lang} and the full set of cross language implications are shown between them. Critically, since $2$ does not consider the possibility of a platypus, her interpretation of $\eta_{mam}$ implicitly excludes the possibility of $\eta_{huev}$ and vice versa; as such, across languages $t_j \imp^{\star} \neg \lambda_{plat}$; since 2 is unaware of platypus, nothing she speaks of refers thereto. But notice the dual statement does not hold $\lambda_{plat} \ {\centernot{\imp}}^{\star}\ \neg t_2 = f_2$. This does not violate \iref{i:neg_cons} exactly because $\lambda_{plat} \ {\centernot{\imp}}^{\star}\ t_2$---it includes things outside of the awareness of agent 2.
\end{example}

\begin{figure}[t]
\centering
\begin{tikzpicture}[
    scale=0.95,
    >=Stealth,
    edge/.style={-{Stealth}, line width=1pt},
    trans/.style={-{>[scale=2.0,length=3.2,width=3.2]}, line width=2pt, lam2},
    node/.style={
        draw,
        rounded corners=3pt,
        minimum height=7mm,
        minimum width=16mm,
        fill=black!5,
        align=center,
        font=\small,
        line width=1pt
    },
]


\begin{scope}[shift={(-4.6,0)}]

\node at (0,6) {\small $\LL_1$};

\node[node] (Fi) at (0,0.0) {$f_1$};

\node[node] (eggonly) at (-2.5,1.6) {$\lambda_{egg}\land \neg\lambda_{mam}$};
\node[node] (mamonly) at ( 2.5,1.6) {$\lambda_{mam}\land \neg\lambda_{egg}$};

\node[node] (plati) at (0,1.6) {$\lambda_{plat}$};
\node[node] (nplati) at (0,3.2) {$\neg \lambda_{plat}$};

\node[node] (eggi) at (-2.4,3.2) {$\lambda_{egg}$};
\node[node] (mami) at ( 2.4,3.2) {$\lambda_{mam}$};

\node[node] (Ti) at (0,4.8) {$t_1$};

\draw[edge] (Fi) -- (plati);
\draw[edge] (Fi) -- (eggonly);
\draw[edge] (Fi) -- (mamonly);

\draw[edge] (eggonly) -- (eggi);
\draw[edge] (mamonly) -- (mami);

\draw[edge] (eggonly) -- (nplati);
\draw[edge] (mamonly) -- (nplati);

\draw[edge] (plati) -- (eggi);
\draw[edge] (plati) -- (mami);

\draw[edge] (eggi) -- (Ti);
\draw[edge] (mami) -- (Ti);
\draw[edge] (nplati) -- (Ti);

\end{scope}

\begin{scope}[shift={(4.6,0)}]

\node at (0,6) {\small $\LL_2$};

\node[node] (Fj) at (0,0.0) {$f_2$};

\node[node] (huevj) at (-1.6,2.4) {$\eta_{huev}$};
\node[node] (mamj)  at ( 1.6,2.5) {$\eta_{mam}$};

\node[node] (Tj) at (0,4.8) {$t_2$};

\draw[edge] (Fj) -- (huevj);
\draw[edge] (Fj) -- (mamj);

\draw[edge] (huevj) -- (Tj);
\draw[edge] (mamj) -- (Tj);

\end{scope}

\draw[trans, {<[scale=2.0,length=4.2,width=4.2]->[scale=2.0,length=4.2,width=4.2]}] (Fi) -- (Fj);
\draw[trans, {<[scale=2.0,length=4.2,width=4.2]->[scale=2.0,length=4.2,width=4.2]}] (mamj) to [out=155,in=50]  (mamonly);
\draw[trans, {<[scale=2.0,length=4.2,width=4.2]->[scale=2.0,length=4.2,width=4.2]}] (huevj) to [out=215,in=-30]  (eggonly);

\draw[trans, {<[scale=2.0,length=4.2,width=4.2]->[scale=2.0,length=4.2,width=4.2]}] (Tj) to [out=155,in=50]  (nplati);

\end{tikzpicture}

\caption{The green arrows define the additional components of $\imp^{\star}$ (the relation itself is the transitive closure of these additional arrows and the within language implication (drawn in black). As a lattice, $\LL_2$ embeds into $\LL_1$.}
\label{fig:cross-lang}

\end{figure}

\begin{theorem}
\label{th:imp-equivalence}
Let $\T$ be a translation over $\<\L_1,\L_2\>$. Letting $i,j \in \{1,2\}$ with $i \neq j$, the following are equivalent:
\begin{enumerate}
\item $\T$ is consistent (satisfies \cref{c:galois}, \cref{c:approximation} and \cref{c:r_dual}), and
\item There exists a unique $\imp^{\star}$ satisfying \iref{i:extensibility}–\iref{i:neg_cons} such that:
for $\lambda_i \in \LL_i$ and $\eta_j \in \LL_j$,
\begin{align}
\label{eq:impstar-def-inner}
\tag{$\star^{-}$}
 \eta_j \imp^\star \lambda_i & \quad \text{ if and only if } \quad \eta_j \imp_j \Tij^{-}(\lambda_i), &&\text{ and}\\
 \label{eq:impstar-def-outer}
\tag{$\star^{+}$}
 \lambda_i \imp^\star \eta_j & \quad \text{ if and only if } \quad  \Tij^{+}(\lambda_i) \imp_j \eta_j
\end{align}
\end{enumerate}
Moreover, if $\L_j$ is complete then 
\begin{equation}
\label{eq:T-from-imp}
\begin{aligned}
\Tij^{-}(\lambda_i) &= \bigvee \{ \eta_j \in \LL_j \mid \eta_j \imp^\star \lambda_i \}, &&\text{ and}\\
\Tij^{+}(\lambda_i) &= \bigwedge\{ \eta_j \in \LL_j \mid \lambda_i \imp^\star \eta_j\},
\end{aligned}
\end{equation}
 for $\lambda_i \in \LL_i$ and where we define $\bigwedge \emptyset = \ast$.

 \gotopf{th:imp-equivalence}
\end{theorem}

\section{Joint State-Spaces}
\label{sec:joint-ss}

\subsection{Joint State Spaces}

A joint state-space is a state-space that can simultaneously make sense of  each agent $i$'s individual understanding as given by $\L_i$. 

\begin{definition}
Call $\<\W,\v_1, \v_2\>$ a \textbf{joint state-space} for $\L_1$ and $\L_2$ if $\W$ is a set  (referred to as a \emph{state-space}) and for $i \in \{1,2\}$,  $\v_i :\LL_i \to 2^W$ is a function such that for all $\lambda_i, \lambda_i' \in \LL_i$,
\begin{itemize}
  \item $\v_i(\lambda_i \land \lambda_i') = \v_i(\lambda_i) \cap \v_i(\lambda_i')$, and
  \item $\v_i(\neg \lambda_i \land \lambda_i') = \v_i(\lambda_i') \setminus \v_i(\lambda_i)$.
\end{itemize}
\end{definition}

Let $\Sigma_{i} = \{\v_i(\lambda) \subseteq \W \mid \lambda \in \LL_i\}$ denote the field of sets that correspond to events in the language $\L$, under the presentation $\v_i$.
A joint state-space preserves implication; in fact, $\<\W,\v_1, \v_2\>$ is a joint state-space if and only if $\lambda_i \mapsto \v_i(\lambda_i)$ defines a Boolean isomorphism between $\LL_i$ and $\Sigma_{i}$. In particular, notice this implies that $\v_i(f_\L) = \emptyset$ and  $\v_i(\lambda_i \lor \lambda_i') = \v_i(\lambda_i) \cup \v_i(\lambda_i')$.
 The idea is that $\v_i$ completely captures the language $\L_i$ over some local part of $\W$, but $\W$ may also include events that are ``unseen'' by $\L_i$.


\subsection{Semantic Translation}

Let $\<\W,\v_1, \v_2\>$ be a joint state-space for $\L_1$ and $\L_2$; for $i \in \{1,2\}$, $\Sigma_i$ is the ring of subsets of $\W$ defined as the image of $\LL_i$.  Let $\Sigma^*_i$ denote $\Sigma_i \cup \{\ast\}$.
Then a \emph{semantic translation operator from $i$ to $j$} is a function $\Vij: \Sigma^*_i \to \Sigma^*_j$. 

A \emph{semantic translation} is then a tuple
$$\V = \<\Vot^{-},\Vot^{+},\Vto^{-},\Vto^{+}\>$$
As with the syntactic translations above, $\Vij^{-}$ and $\Vji^{-}$ are referred to as \emph{inner} translation operators and $\Vij^{+}$ and $\Vji^{+}$ as \emph{outer} translation operators. We require that each operator preserves $\ast$ so that $\Vij(\ast) = \ast$.

Call a semantic translation, the $\<\Sigma_1, \Sigma_2\>$ \emph{approximation} if
\begin{align*}
  \Vij^{-}(E) &= \bigcup \Big\{F \in \Sigma_{j} \mid F \subseteq E \Big\} && \text{and} \\
  \Vij^{+}(E) &= \bigcap \Big\{F \in \Sigma_{j} \mid E \subseteq F\Big\} 
\end{align*}
where we define $\bigcap \emptyset = \ast$.




\begin{proposition}
\label{prop:semantic-to-trans} 
Let $\<\L_1,\L_2\>$ be complete and $\T$ a consistent translation between them, with $\imp^\star$ the unique relation satisfying \eqref{eq:T-from-imp}. 
 Let $\<\W,\v_1,\v_2\>$ be a joint state-space and $\V$ the $\<\Sigma_{1},\Sigma_{2}\>$ approximation.
The following are equivalent:
\begin{enumerate}
  \item $\eta \imp^\star \lambda$ if and only if $\v_j(\eta) \subseteq \v_i(\lambda)$,
  \item $\V = \v_j \circ \T \circ \v_i^{-1}$, and
    \item $\T = \v_j^{-1} \circ \V \circ \v_i$.
\end{enumerate}
 where this notation means for each operator $\Tij$, to pre-compose with $\v_i^{-1}$ and post-compose $\v_j$ and vice versa.

\gotopf{prop:semantic-to-trans} 
\end{proposition}

\subsection{Existence of a Joint State-Space}
\label{sec:existence}

Proposition \ref{prop:semantic-to-trans} essentially states that, when a joint state-space exists for $\<\L_1,\L_2\>$, then the three different perspectives on translation outlined by the proceeding three sections---syntactic translation operators $\T$, cross language implication $\imp^\star$, and semantic translation $\V$---all carry the same information about the relation between the two languages; that is, Proposition \ref{prop:semantic-to-trans} provides a recipe for losslessly switching from one description of the problem into another. It does not, however, ensure that a joint state-space exists in the first place. Worry not, however, as this is guaranteed by our main representation result

\begin{theorem}
\label{thm:state-space-existance} 
Let $\<\L_1,\L_2\>$ be complete and $\T$ a translation between them. Then,
the following are equivalent:
\begin{enumerate}
  \item \label{thm:sse:T}$\T$ satisfies \cref{c:galois}---\cref{c:r_dual},
  \item \label{thm:sse:i} The relation $\imp^\star$ defined by \eqref{eq:T-from-imp} satisfies \iref{i:extensibility}---\iref{i:neg_cons}
    \item \label{thm:sse:state-space} There exists a joint state-space $\<\W, \v_1, \v_2\>$ such that $\v_j \circ \T \circ \v_i^{-1}$ is the $\<\Sigma_{1},\Sigma_{2}\>$ approximation.
\end{enumerate}
\gotopf{thm:state-space-existance}
\end{theorem}

\section{Common Languages and Joint Languages}
\label{sec:common-ss}

In this section we consider the idea of a common language, which can express those things that are expressible by both agents, and a join languages, which can express anything that is expressible by either agent.

\subsection{Common Languages}
A common language collects those propositions that can be translated perfectly, and therefore, discussed unambiguously. To model this, we define an abstract notion of a `common language' that can be simultaneously embedded into both $\L_1$ and $\L_2$. These embeddings are \emph{relative} to agents' awareness, and so, negation will not be generally be preserved. Instead, the common language preserves only relative negation.

\begin{definition}
\label{definition-common-language}
Call $(\L, \epsilon_1, \epsilon_2)$ a \textbf{common language} for $\L_1$ and $\L_2$ if $\L = (L,B)$ is a language, and for $i \in \{1,2\}$,  $\epsilon_i :\LL \to \LL_i$ such that for all $\lambda, \lambda' \in \LL$
\begin{itemize}
  \item $\epsilon_i(\lambda\land \lambda') = \epsilon_i(\lambda) \land \epsilon_i(\lambda')$,
  \item $\epsilon_i(\neg \lambda \land \lambda') = \neg \epsilon_i(\lambda_i) \land \epsilon_i(\lambda_i')$.
\end{itemize}
\end{definition}

It follows from the first condition that logical implication of the common language is preserved, and so, each individual language contains (an image of) the common language.

\begin{remark}
\label{lemma-common-lang-implication} If $\L$ is a common language for $L_1$ and $\L_2$, then for all $\lambda, \eta \in \LL$: 
\begin{equation*}
\lambda \imp_{\L} \eta \quad \text{ if and only if }\quad \epsilon_i(\lambda) \imp_{i} \epsilon_i(\eta)
\end{equation*}%
for $i \in \{1,2\}$. As such, $\epsilon_i$ is injective.
\end{remark}

From a consistent translation $\T$, we can construct a common language by examining the set of perfect translations. Indeed we have the following

\begin{proposition}
\label{lemma-common-lang-exist} Let $\<\L_1,\L_2\>$ be complete and  $\T$ be a consistent translation between them, with $\imp^\star$ the unique relation satisfying \eqref{eq:T-from-imp}, then the sets
\begin{enumerate}
  \item $\{\lambda \in \LL_i \mid \Tij^{-}(\lambda) = \Tij^{+}(\lambda)\}$,
  \item $\{\lambda \in \LL_i \mid \lambda \Leftrightarrow^\ast \eta, \text{ for some } \eta \in \LL_j\}$,
   \item $\{\lambda \in \LL_i \mid \Tji^{+}(\Tij^{+}(\lambda)) = \lambda\}$, and
   \item $\{\lambda \in \LL_i \mid \Tji^{-}(\Tij^{-}(\lambda)) = \lambda\}$
\end{enumerate}
coincide and, so long as they contain more that one proposition, serve as a common language for $\<\L_1,\L_2\>$.
\end{proposition}
\gotopf{lemma-common-lang-exist}

\noindent Since $f_1 \Leftrightarrow^\ast f_2$ by \iref{i:explosion}, Proposition \ref{lemma-common-lang-exist} establishes that a common language exists so long as there is any other perfect translation.

It is well known that the set of fixed points of a Galois connection between lattices themselves form a lattice (that is complete whenever the original lattices are). Moreover, the restriction of the Galois connection to these fixed points is a lattice isomorphism. It is easy to see that whenever $\lambda$ can be translated perfectly, then it is a fixed point:

\begin{remark}
\label{lemma-fixed-points} If $\lambda \in \LL_i $ is such that $\Tij^{-}(\lambda) = \Tij^{+}(\lambda)$, then $\lambda$ is a fixed point of translation: $\Tji^{+}(\Tij^{-}(\lambda)) = \Tji^{-}(\Tij^{+}(\lambda)) = \lambda$.
\end{remark}


One might hope, then, that these fixed points define a common language. This however, is not the case, as shown by the following counter-example.

\begin{example}
  Let $\LL_1$ be the Boolean algebra with two atoms (labeled $\lambda$ and $\neg\lambda$) and $\LL_2$ be the Boolean algebra with three atoms (labeled $\eta_a$, $\eta_b$, and $\eta_c$). Then let the translations be as follows: $t_1$ and $t_2$ and $f_1$ and $f_2$ are perfect translations of each other, $\Tot$ is defined by
      \begin{align*}
        \Tot^{+}(\lambda) = \eta_b \lor \eta_c 
        && \Tot^{-}(\lambda) = \eta_c 
        &&\Tot^{+}(\neg \lambda) = \eta_a \lor \eta_b 
        && \Tot^{-}(\neg \lambda) = \eta_a
    \end{align*}
and $\Tto$ is defined by
      \begin{align*}
        \Tto^{+}(\eta_a) &= \neg \lambda 
        && \Tto^{-}(\eta_a) = f_i
        &&\Tto^{+}(\eta_a \lor \eta_b ) = t_i
        && \Tto^{-}(\eta_a \lor \eta_b ) = \neg\lambda \\
        \Tto^{+}(\eta_b) &= t_i
        && \Tto^{-}(\eta_b) = f_i
        &&\Tto^{+}(\eta_a \lor \eta_c ) = t_i
        && \Tto^{-}(\eta_a \lor \eta_c ) = f_i \\
        \Tto^{+}(\eta_c) &= \lambda
        && \Tto^{-}(\eta_c) = f_i
        &&\Tto^{+}(\eta_b \lor \eta_c ) = t_i
        && \Tto^{-}(\eta_b \lor \eta_c ) = \lambda
    \end{align*}
    Here we have that $\lambda$ is a fixed point: $\Tto^{+}(\Tot^{-}(\lambda)) = \Tto^{-}(\Tot^{+}(\lambda))  = \lambda$. However, $\lambda$ is not a perfect translation $\Tot^{+}(\lambda) = \eta_b \lor \eta_c \neq \eta_c = \Tot^{-}(\lambda)$.
\end{example}

\begin{figure}[]
\centering
\begin{tikzpicture}[
  scale=1.0,
  >=Stealth,
  edge/.style={-{>[scale=2.3,length=1.4,width=1.4]}, line width=1pt},
  trans/.style={-{>[scale=2.0,length=3.2,width=3.2]}, line width=1.5pt, dashed},
  node/.style={
    draw,
    rounded corners=3pt,
    minimum height=7mm,
    minimum width=8mm,
    fill=black!5,
    align=center,
    font=\small,
    line width=1pt
  },
  dots/.style={font=\Large}
]

\begin{scope}[shift={(-3.2,0)}]

  \node[node] (LF)  at (0,1.2) {$f_1$};
  \node[node] (LB)  at (-1.2,2.7) {$\neg\lambda$}; 
  \node[node] (LA) at ( 1.2,2.7) {$\lambda$};

  \node[node] (LT)  at (0,4.2) {$t_1$};

  \draw[edge] (LF) -- (LB);
  \draw[edge] (LF) -- (LA);

  \draw[edge] (LA) -- (LT);
  \draw[edge] (LB) -- (LT);

\end{scope}

\begin{scope}[shift={(3.2,0)}]

  \node[node] (RF)  at (0,1) {$f_2$};

  \node[node] (RA)  at (1.8,2.2) {$\eta_a$};
  \node[node] (RB) at ( 0,2.2) {$\eta_b$};
    \node[node] (RC)  at (-1.8,2.2) {$\eta_c$};

      \node[node] (RAB)  at (1.8,3.4) {$\eta_a \lor \eta_b$};
  \node[node] (RAC) at ( 0,3.4) {$\eta_a \lor \eta_c$};
    \node[node] (RBC)  at (-1.8,3.4) {$\eta_b \lor \eta_c$};

  \node[node] (RT)  at (0,4.6) {$t_2$};

  \draw[edge] (RF) -- (RA);
  \draw[edge] (RF) -- (RB);
  \draw[edge] (RF) -- (RC);

  \draw[edge] (RAB) -- (RT);
  \draw[edge] (RAC) -- (RT);
  \draw[edge] (RBC) -- (RT);

  \draw[edge] (RA) -- (RAB);
  \draw[edge] (RA) -- (RAC);
  \draw[edge] (RB) -- (RAB);
  \draw[edge] (RB) -- (RBC);
  \draw[edge] (RC) -- (RAC);
  \draw[edge] (RC) -- (RBC);

\end{scope}


\draw[trans,  {<[scale=1.5,length=4.2,width=4.2]->[scale=1.5,length=4.2,width=4.2]}, solid, color=lam1] (LT) -- (RT);
\draw[trans, {<[scale=1.5,length=4.2,width=4.2]->[scale=1.5,length=4.2,width=4.2]}, solid, color=lam1] (LF) -- (RF);


\draw[trans, {<[scale=1.5,length=4.2,width=4.2]->[scale=1.5,length=4.2,width=4.2]}, color=lam1] (LA) -- (RC);
\draw[trans, {<[scale=1.5,length=4.2,width=4.2]->[scale=1.5,length=4.2,width=4.2]}, color=lam1] (LA) -- (RBC);

\draw[trans, color=lam1] (RBC) -- (LT);
\draw[trans, color=lam1] (RC) -- (LF);


\end{tikzpicture}
\caption{$\lambda$ is a fixed point but not a perfect translation.}
\label{fig:fixedNotPerfect}
\end{figure}

\subsection{Joint Languages}
A joint language can be thought of as a `universal' language that contains statements that correspond to each agent's individual language, and thus, in general, the joint language is beyond the comprehension of either agent individually. A joint language is a language for a joint state-space. To model this, we define an abstract notion of a `joint language' as a language that both $\L_1$ and $\L_2$ can be embedded into. These embeddings are \emph{relative} to what the agents are aware of and so negation will not be generally be fully preserved in cases of restricted awareness. 

\begin{definition}
\label{definition-joint-language}
Call $(\L, \pi_1, \pi_2)$ a \textbf{joint language} for $\L_1$ and $\L_2$ if $\L = (L,B)$ is a language, and for $i \in \{1,2\}$,  $\pi_i :\LL_i \to \LL$ is a such for all $\lambda, \lambda' \in \LL_i$
\begin{itemize}
  \item $\pi_i(\lambda_i \land \lambda_i') = \pi_i(\lambda_i) \land \pi_i(\lambda_i')$,
  \item $\pi_i(\neg \lambda_i \land \lambda_i') = \neg \pi_i(\lambda_i) \land \pi_i(\lambda_i')$.
\end{itemize}
\end{definition}

It follows from the first condition that logical implication in each of the individual languages is exactly preserved, and therefore, the joint language must contain (images of) each of the individual languages.

\begin{remark}
\label{lemma-joint-lang-implication} If $\L$ is a joint language for $%
\L_1$ and $\L_2$, then for all $\lambda_i, \eta_i \in \LL_i$: 
\begin{equation*}
\lambda_i \imp_{i} \eta_i \quad \text{ if and only if }\quad \pi_i(\lambda_i) \imp_{\L} \pi_i(\eta_i)
\end{equation*}%
for $i \in \{1,2\}$. As such, $\pi_i$ is injective.
\end{remark}

Notice that the maps $\pi_1$ and $\pi_2$ contain all the information about how
the two languages relate to each other. The structure of these embeddings
(and indeed their existence) crucially depends restrictions of the joint language $\L = (L,B)$ (i.e., depends on the elements $B$ of the language that are required to be true); these axioms tacitly relate the
meaning of the propositions in the two languages.

In an immediate Corollary to Theorem \ref{thm:state-space-existance} and Remark \ref{lemma-algebra-to-lang}, a joint language for $\<\L_1,\L_2\>$ exists exactly when a consistent translation exists between them.

\begin{figure}[]
\centering
\begin{tikzpicture}[
  arr/.style={-{Stealth[length=6pt]}, thick},
  yscale=.8,
]
 
\node (B) at ( 0,   0  ) {$\LCom$};
\node (L) at (-3,   3  ) {$\L_1$};
\node (R) at ( 3,   3  ) {$\L_2$};
\node (T) at ( 0,   6  ) {$\LJnt$};
 
\draw[arr] (B) -- node[left]  {$\epsilon_1$} (L);
\draw[arr] (B) -- node[right] {$\epsilon_2$} (R);
\draw[arr] (L) -- node[left]  {$\pi_1$}      (T);
\draw[arr] (R) -- node[right] {$\pi_2$}      (T);
 
\draw[arr] (L) to[bend left=20]
           node[above] {$\Tot$} (R);
 
\draw[arr] (R) to[bend left=20]
           node[below] {$\Tto$} (L);
 
\end{tikzpicture}
\caption{A communicative diagram of the embedding relationship between the common and joint languages.}
\label{fig:embed}
\end{figure}

The common language can be embedded into the joint language, with either individual language sitting in between. Indeed, let $\<\L_1,\L_2\>$ be complete and  $\T$ a consistent translation between them, $\<\LCom, \epsilon_1,\epsilon_2\>$ be  the common language that is constructed by the set of perfect translations (Proposition \ref{lemma-common-lang-exist}) and $\<\LJnt, \pi_1,\pi_2\>$ be the joint language that corresponds to the joint state-space constructed from $\T$ (Theorem \ref{thm:state-space-existance}). Then $\LCom$ embeds into $\L^{JNT}$ via $\pi_i \circ \epsilon_i$; of course, this embedding does not depend on the choice of  $i\in \{1,2\}$. Further, for $i \neq j$, the maps $\pi_i \circ \Tij \circ \epsilon_i$ produce the same embedding (for $\Tij^{-}$ or $\Tij^{+}$). These relations are shown in Figure \ref{fig:embed}.

The relationship between the common language and the joint language, of which it is a sublattice, sets bounds on the capacity of agents to communicate. Communication in the common language is unambiguous. Statements in the joint language are comprehensible to both agents but may be interpreted differently. 


\subsection{Comparative Awareness}

In this section, we show how notions of comparative awareness can equivalently be defined directly using a joint state-space (as is standard) or a joint language.
Let $\L$ be a joint language for $\L_i$ and $\L_j$, with tautology and contradiction given by $t$ and $f$, and implication $\imp_{\L}$.

Coarsening arises when an agent fails to distinguish between events that are in fact different. Say that $\L_i$ is a \emph{pure coarsening} of $\L_j$ if $\pi_i(t_i)=\pi_j(t_j)=t$ and for all $\lambda_i \in \LL_i$, there is an $\eta_j \in \LL_j$ such that $\pi_i(\lambda_i) \Leftrightarrow_{\L} \pi_j(\eta_j)$. Thus, every proposition in $\L_i$ can be expressed by a proposition in $\L_j$, but not necessarily vice versa. In semantic terms, the languages agree on the universal event: $\v_i(t_i) = \v_j(t_j) = W(\L)$ but the algebra of events generated $\L_i$ is coarser than that generated by $\L_j$: $\Sigma_{i} \subseteq \Sigma_{j}$.

Restriction, on the other hand, refers to a situation in which an agent is unaware of events that are in fact possible. Say that $\L_i$ is a \emph{pure restriction} of $\L_j$ if $\pi_i(t_i)\imp_{\L} \pi_j(t_j)$ and $\pi_i \circ\pi_j^{-1}$ is a Boolean isomorphism between $\LL_i$ and $\{\lambda \in \LL_j\mid \pi_j(\lambda_j)\imp_{\L} \pi_i(t_i)\}$. Every proposition considered possible in $\L_i$ is thus considered possible in $\L_j$, but not necessarily vice versa. In semantic terms, $\Sigma_{i} = \{E \cap \v_i(t_i)\mid E \in \Sigma_{j}\}$: $i$ considers possible a subset of the \emph{states} considered by $j$, but over this subset their awareness agrees.

Finally, say that $\L_i$ is \emph{less aware} than $\L_j$ if there is a language $\L^\dag$ such that $\L_i$ is a pure coarsening of $\L^\dag$ and $\L^\dag$ is a pure restriction of $\L_j$. Clearly, if $\L$ is a joint language of $\L_i$ and $\L_j$, then each $\L_i$ and $\L_j$ are combinations less aware than $\L$. In particular, for language $\L_i$, the set $A_i=\v_i(t_i)$ determines the restriction, whereas the coarsening is captured by $\Sigma_{i}$, which defines the expressible subsets of $A_i$.

Note that in both cases of pure coarsening and pure restriction, we have that $\Sigma_{i} \subseteq \Sigma_{j}$. It is easy to see that the reverse is also true: if $\Sigma_{i} \subseteq \Sigma_{j}$, then $\L_i$ is a combination of coarsening and restriction of $\L_j$. Extending this line of reasoning, the partial order over languages defined by ``$\L_i$ is less aware than $\L_j$'' can in fact be characterized through the relation of $\L_j$ being a joint language for the pair, as shown by the following.

\begin{remark}
\label{corollary-comparative-unawareness-1}
Consider two languages $\L_j$ and $\L_i$. Then the following are equivalent:

\begin{enumerate}
  \item $\L_i$ is less aware than $\L_j$ 
  \item $\L_j$ serves as a minimal joint language for the pair
  \item $\L_i$ serves as a common language for the pair
  \item There exists a joint state-space such that $\Sigma_{i} \subseteq \Sigma_{j}$
  \item $\Tij^{+}(\lambda) = \Tij^{-}(\lambda)$ for all $\lambda \in \LL_i$.
\end{enumerate}
\end{remark}

For example, consider the case where of $\L_i = (L_i,B_i)$ and $\L_j = (L_j,B_j)$ where $L_i \subseteq L_j$ and $B_j \subseteq B_i$. Then Remark \ref{corollary-comparative-unawareness-1} establishes that $\LL_i$ serves as a joint language. In particular, we have that if $B_i = B_j$ then $\L_i$ is a pure coarsening of $\L_j$, whereas if $L_i = L_j$, then $\L_i$ is a pure restriction.

\section{Literature}
\label{sec:lit}

The standard decision-theoretic model of uncertainty, both for individual decisions and for games, assumes a commonly-known state space (Savage, 1954). This assumption is maintained in generalizations of the model, such as rank-dependent utility \citep{quiggin1982theory}, cumulative prospect theory \citep{tversky1992advances}, and choice under ambiguity \citep{gilboa1989maxmin,klibanoff2005smooth}. \cite{tversky1994support} are among the first to note that the proposition used to refer to an event may bias its perception. They discuss the impact of such misperception on probabilistic judgments. \cite{minardi2019subjective} use the change in preferences due to updating to derive a subjective state space, which may differ from the objective one and also discuss implications for probabilistic judgments. \cite{halpern2003reasoning} describes the states of the world" approach as semantic, distinguishing this from syntactic, or propositional representations. The agents' different views of the world are naturally expressed by a language and its propositions. \cite{halpern2015syntax} points out the advantages of the syntactic approach, which is also defended by \cite{feinberg2000characterizing}. \cite{blume2021constructive} axiomatize expected utility maximization whereby the state-space is endogenously derived from preferences over programs described in terms of syntactic propositions. Their framework allows for misperception of the objective'' state space as described by  \cite{tversky1994support}.

The relationship between syntax and semantics becomes more complex under bounded awareness. In the case of a single agent, the primary problem is to compare the model of the world held by the agent with that which they might hold if they were fully aware. A crucial distinction is that between coarse awareness and restricted awareness.\footnote{This distinction has been pointed out by a number of authors, beginning with \cite[pp. 87-94]{carnap1971basic}, but has not been fully appreciated in the literature on unawareness} It is most naturally expressed in syntactic terms: an agent displays coarse awareness if their language does not include a sufficiently rich set of propositions to distinguish between two (or more) possibilities. Restricted awareness arises when an agent ignores some possibility, implicitly assuming propositions expressing that possibility to be false.\footnote{An important implication of restricted awareness is that agents' probability assignments for some propositions must be incorrect. By contrast, coarse awareness does not, in itself, imply incorrect probability judgements. We address this issue in other work.}

Multiple works have demonstrated the relevance of differential awareness in economic applications. \cite{heifetz2006interactive} address the issue of interactions between agents with coarse awareness. In syntactic terms, coarse awareness is represented by the idea that each agent has access to a language derived from a subset of the propositions in a descriptively complete language sufficiently rich to discriminate between all relevant possibilities. Common awareness is generated by the intersection of the individual languages. \cite{HEIFETZ2008304} and \cite{halpern2009reasoning} provide axiomatizations for this model. Although their model introduces a mapping (projection) from the events of which one agent is aware to those of which another agent is aware, these mappings are only defined for agents whose awareness is comparable in the sense of coarsening / refinement, but not in the general case.

\cite{hill2006} considers distinct languages
(models of the world) entertained by the same agent but at different
instances,
interpreting these as representations of distinct states of awareness. He relates such
languages by two binary relations: one stating that two propositions from
distinct languages are identical; a second one stating that propositions
from distinct languages are logically equivalent. He defines the ``fusion''
operation between two such languages by taking the cross-product of the
corresponding Boolean algebras and then its quotient with respect to the
so-defined equivalence relations. Our model is more general in that we
account not only for equivalences, but also for implications. As a result,
we can capture a richer set of dependencies between the propositions of the
two languages (e.g., non-nested partitions of a joint state-space). Our
construction of the joint language and the joint state-space is thus more
complex and based on identifying and combining the ultrafilters of the two
Boolean algebras. 

Although their formal model is one of coarse awareness, \cite{heifetz2006interactive} consider a market interaction between two agents each with restricted awareness of propositions relevant to the market value of a firm. In this example, bounded awareness leads to willingness to trade, even though a no-trade result would hold under full awareness. Similar problems of differential restricted analysis are considered by \cite{gabaix2006shrouded} (modelling competitive markets), \cite{filiz2012incorporating} (contracts), and \cite{li2014vertically} (duopoly). All of these papers use a semantic framework, although propositional labels may be attached to states to enhance intuition. \cite{piermont2017introspective} demonstrates that bounded and differential awareness may lead to preferences for flexibility, as well as to strategic concealment of actions, which in turn gives rise to incomplete contracts.

The issue of translation between different languages is to our knowledge first introduced by \cite[p. 224]{carnap1937logical}: he studies mappings which preserve logical implication and examines the preservation of other logical operations. In our setting, his analysis applies to the case of coarsening.\footnote{\cite{carnap1937logical} uses the term sublanguage'' to refer to a language which is less expressive. Unawareness in terms of restriction corresponds to a sublanguage of the fourth type" in \cite{carnap1971basic}.} For the case of restriction, \cite[pp. 93-97]{carnap1971basic} states that translation does not preserve logical equivalence and notes the challenge of dealing with contradictions, which might emerge based on empirically observable facts.

When agents' languages differ in expressiveness, they are represented by distinct sub-algebras of the joint state space. Translation consists in finding the best approximation of the event expressed by a proposition, which may not be measurable in the algebra of the target language. This is formally similar to the problem of learning based on imprecise evidence, as in \cite{dempster1967upper} and \cite{shafer1976mathematical}. The upper and lower approximations of an event give rise to the outer and inner translation operator in our model, similarly to the inner and outer measures in the theory of imprecise probabilities. It is well-known that the inner and outer measures satisfy a duality property: specifying one of them uniquely pinpoints the other. Under common awareness, a similar duality exists between the notions of knowledge / belief and possibility in the theory of epistemic knowledge. In contrast, as already observed by \cite{modica1999unawareness}, duality does not hold for the inner and outer translation when awareness takes the form of restriction. They suggest a restricted notion of duality conditional on awareness. In the framework of \cite{heifetz2006interactive},  \cite{fukuda2023strategic} shows that with this notion of duality, and given an awareness operator, there is a unique correspondence between belief and possibility. This is also the case in our model, whenever the awareness of the two agents takes the form of coarsening. 
For this case, translation preserves tautologies, and our axiom \cref{c:r_dual},
Restricted Duality, is mathematically equivalent to the duality condition
introduced in \cite{fukuda2023strategic}. When unawareness takes the form of
restriction, this equivalence fails. \cref{c:r_dual} thus has to take into account the
fact that negations of $\Leftrightarrow^{\ast }$-equivalent propositions
need not be $\Leftrightarrow^{\ast }$-equivalent.%

In an economic context, translation is discussed by \cite{viero2023lost} who is interested in how an external observer can recognize the fact that a decision maker cannot distinguish two contingencies. \cite{grant2018contracting} consider the case of contracting where both parties are fully aware, but use different languages. Although the contractual language is common, the two parties may disagree about its interpretation, which may result in contractual disputes. Both \cite{piermont2021hypothetical} and \cite{kops2026choice} introduce an interpretation functional between an ``objective'' and a ``subjective'' state space, which is similar to our translation operator, but does not involve translation across different subjective state spaces. \cite{piermont2021unforeseen} shows how an agent can incorporate new statements/events into their subjective state-space using a mapping between the objective and the subjective state space. In the terminology of the current paper, the probability of the novel event is restricted to reside between the probability of the inner and outer translations.

\section{Concluding Comments}

Beginning with Savage's discussion of breaking eggs to make an omelette, the
predominant mode of analysis in decision theory has been to exploit the
formal structure of state spaces to represent beliefs and preferences, while
motivating the analysis with propositional labels for states and events. The
lattice structure of propositional algebras has been relied on implicitly
but not used in formal analysis. For the case of individual decisions with
fixed awareness, Stone's representation theorem shows that the two
representations are isomorphic.

When we consider interactions between agents with different awareness, or an
individual with changing awareness over time, the advantages of a syntactic
approach come to the fore. The notion of translation between languages is
more natural and tractable than the alternative of mappings between state
spaces. 

In this paper, we have derived the properties under which well-behaved
translations between languages exist, and analysed the properties of joint
and common languages. This analysis yields an equivalent representation in
the more familiar framework of state spaces.

In future work, we will extend this analysis to encompass beliefs and belief
revision, with potential applications to finance markets, as foreshadowed in
the introduction.

Also, because restricted awareness impedes the classical duality of negation (i.e.,, negations of $\Leftrightarrow^{\ast }$-equivalent propositions
need not be $\Leftrightarrow^{\ast }$-equivalent, leading to the weaker duality embodied by \cref{c:r_dual} and \iref{i:neg_cons}), we believe that it is fruitful to examine awareness and translation starting with non-Boolean languages that capture weaker notions of negation. For example, we hope to extend these results to the case where the agent's individual languages are Heyting algebras or even distributive lattices; note that a version of Theorem \ref{th:imp-equivalence} holds where both \cref{c:r_dual} and \iref{i:neg_cons} are dropped. 

\newpage

\section{Proofs}

\subsection{Lemmas}

\begin{lemma}
\label{lem:gal-implications} Let $\T$ satisfy \hyperref[c:galois]{%
\textcolor{lam1}{\textup{\textbf{C\ref{c:galois}}}}} and \hyperref[c:approximation]%
{\textcolor{lam1}{\textup{\textbf{C\ref{c:approximation}}}}}. Let $\lambda
\in \mathcal{L}^{\ast}_i$. The following hold:

\begin{enumerate}
[label=(\roman*)]

\item \label{lem:imps:closure} $\lambda \Rightarrow_i \T_{j \to
i}^{-}(\T_{i \to j}^{+}(\lambda))$

\item \label{lem:imps:interior}$\T_{j \to i}^{+}(\T_{i \to
j}^{-}(\lambda)) \Rightarrow_i \lambda$

\end{enumerate}
\end{lemma}

\begin{proof} Let $\lambda \in \LS_i$:
\begin{enumerate}[label=(\roman*)]
  \item By reflexivity we have $\Tij^{+}(\lambda) \imp_j \Tij^{+}(\lambda)$; applying \eqref{eq:galois} delivers the relation.
  \item By reflexivity we have $\Tij^{-}(\lambda) \imp_j \Tij^{-}(\lambda)$; applying \eqref{eq:galois} delivers the relation.
  \qedhere
\end{enumerate}
\end{proof}

\begin{lemma}
\label{lem:selfimply} 
Assume $\L_j$ is complete.
Let $\imp^\star$ satisfy \iref{i:extensibility}--\iref{i:explosion}. Let $\lambda \in \LL_i$, then 
\begin{equation} 
\label{eq:selfimpIn}
\bigvee \{ \eta \in \LL_j \mid \eta \imp^\star
\lambda \} \imp^\star \lambda.
\end{equation}
Moreover, if there exists some $\eta \in \LL_j$ such that $\lambda \imp^\star \eta$, then also 
\begin{equation} 
\label{eq:selfimpOut}
\lambda \imp^\star \bigwedge \{ \eta \in \LL_j \mid
\lambda \imp^\star \eta \}.
\end{equation}
\end{lemma}

\begin{proof}
By \iref{i:explosion}, $\bigvee \{ \eta \in \LL_j \mid \eta \imp^\star \lambda \}$ and $\bigwedge \{ \eta \in \LL_j \mid \lambda \imp^\star \eta \}$ are non-empty; \eqref{eq:selfimpIn} then follows immediately from \iref{i:connective} and completeness.
\end{proof}

\begin{pproof}{prop:equivalence} We will show each property:
    \begin{itemize}
      \item[\tref{t:contradiction}:] By definition $f_{i}$ we have $f_{i} \imp_i \Tji^{-}(f_{i})$. Applying \eqref{eq:galois} we have $\Tij^{+}(f_{i}) \imp_j f_{j}$; it follows that $\Tij^{+}(f_{i}) = f_{j}$. Further, by \cref{c:approximation}, $\Tij^{-}(f_{i}) \imp_i f_{j}$, or that $\Tij^{-}(f_{i}) = f_{j}$, as well. 
      
      \item[\tref{t:concreteness}:] Let $\Tij^{-}(\lambda) = \ast$. By Lemma \ref{lem:gal-implications}\ref{lem:imps:interior}, $\Tji^{+}(\ast) \imp_i \lambda$; thus since $\Tji^{+}$ preserves $\ast$, $\ast \imp_i \lambda$, or $\lambda = \ast$.
      
      \item[\tref{t:monotonicity}:] Let $\lambda \imp_i \lambda'$. By Lemma \ref{lem:gal-implications}\ref{lem:imps:closure}, $\lambda \imp_i  \Tji^{-}(\Tij^{+}(\lambda'))$. Applying \eqref{eq:galois} delivers the relation.
      
      \item[\tref{t:conjunction}:] On account of \tref{t:monotonicity}, it suffices to show that $\Tij^{-}(\lambda) \land \Tij^{-}(\lambda) \imp_j  \Tij^{-}(\lambda \land \lambda')$. From Lemma \ref{lem:gal-implications}\ref{lem:imps:interior}, we have $\Tji^{+}(\Tij^{-}(\lambda)) \imp_i \lambda$. So, by \tref{t:monotonicity}, $\Tji^{+}(\Tij^{-}(\lambda) \land \Tij^{-}(\lambda')) \imp_i \lambda$; likewise $\Tji^{+}(\Tij^{-}(\lambda) \land \Tij^{-}(\lambda')) \imp_i \lambda'$, and so $\Tji^{+}(\Tij^{-}(\lambda) \land \Tij^{-}(\lambda')) \imp_i \lambda \land \lambda'$. Applying \eqref{eq:galois} delivers the relation.
      
      \item[\tref{t:disjunction}:] Analogous to the proof of  \tref{t:conjunction}. \qedhere
    \end{itemize}
\end{pproof}

\begin{tproof}{th:imp-equivalence}
That (2) implies (1) simply requires checking each condition. 
\begin{itemize}
  \item[\cref{c:galois}:] Fix $\eta_j \in \LS_j$ and $\lambda_i \in \LS_i$. Let $\eta_j \imp_j \Tij^{-}(\lambda_i)$. By the reflexivity of $\imp_j$, we have $\Tij^{-}(\lambda_i) \imp_j \Tij^{-}(\lambda_i)$, and so by  \eqref{eq:impstar-def-inner},
  $\Tij^{-}(\lambda_i) \imp^\star \lambda_i$. It then follows from \iref{i:extensibility} and \iref{i:transitivity} that $\eta_j \imp^\star \lambda_i$. As such, $\Tji^{+}(\eta_j) \imp_i \lambda_i$ by \eqref{eq:impstar-def-outer}.
  The other direction is analogous. 



  \item[\cref{c:approximation}:] If $\Tij^{+}(\lambda_i) = \ast$, the result is immediate, so assume it is not.
 By the reflexivity of $\imp_j$, we have $\Tij^{-}(\lambda_i) \imp_j \Tij^{-}(\lambda_i)$ and $\Tij^{+}(\lambda_i) \imp_j \Tij^{+}(\lambda_i)$. Applying \eqref{eq:impstar-def-inner} to the former and  \eqref{eq:impstar-def-outer} to the later, yields $\Tij^{-}(\lambda_i) \imp^\star \lambda_i$ and $\lambda_i  \imp^\star \Tij^{+}(\lambda_i)$, respectively. The result follows from \iref{i:transitivity} and \iref{i:extensibility}.

\item[\cref{c:r_dual}.] 

Let  $\lambda_i \in \LL_i$ such that $\Tij^{+}(\lambda_i) \neq \ast_j$. Thus, $\Tij^{+}(\lambda_i) \imp_j t_j$, or by \eqref{eq:impstar-def-outer}, $\lambda_i \imp^\star t_j$.
We have, 
\begin{align*}
  &\Tij^{-}(\neg \lambda_i) \imp^\star \neg \lambda_i && \text{(from \eqref{eq:impstar-def-inner})} \\
\text{iff} \quad &\lambda_i \imp^\star \neg \Tij^{-}(\neg \lambda_i)  && \text{(from \iref{i:neg_cons})} \\
\text{iff} \quad &\Tij^{+}(\lambda_i )\imp_j \neg \Tij^{-}(\neg \lambda_i)  && \text{(from \eqref{eq:impstar-def-outer})} \\
\text{iff} \quad & \Tij^{-}(\neg \lambda_i) \imp_j \neg \Tij^{+}(\lambda_i )  && \text{(from Boolean negation in $\LL_j$)} 
\end{align*}
Since also $\Tij^{-}(\neg \lambda_i) \imp_j \Tij^{-}(t_i)$ from \tref{t:monotonicity}, we conclude that $\Tij^{-}(\neg \lambda_i) \imp_j \neg \Tij^{+}(\lambda_i) \land \Tij^{-}(t_i)$.

In the other direction, note that $\neg \Tij^{+}(\lambda_i) \land \Tij^{-}(t_i) \imp^\star \Tij^{-}(t_i) \imp^\star t_i$, where the later implication arises from \eqref{eq:impstar-def-inner}).
We have, 
\begin{align*}
  &\lambda_i \imp^\star \Tij^{+}(\lambda_i ) && \text{(from \eqref{eq:impstar-def-outer})} \\
\text{iff} \quad &\lambda_i \imp^\star \Tij^{+}(\lambda_i ) \lor  \neg \Tij^{-}(t_i)  && \text{($\lor$-introduction)} \\
\text{iff} \quad &\lambda_i \imp^\star \neg(\neg \Tij^{+}(\lambda_i ) \land \Tij^{-}(t_i))  && \text{(De Morgan's law)} \\
\text{iff} \quad & \neg \Tij^{+}(\lambda_i ) \land \Tij^{-}(t_i) \imp^\star \neg \lambda_i  && \text{(from \iref{i:neg_cons})} \\
\text{iff} \quad & \neg \Tij^{+}(\lambda_i ) \land \Tij^{-}(t_i) \imp_j \neg \Tij^{-}(\lambda_i)  && \text{(from \eqref{eq:impstar-def-inner})} 
\end{align*}
and so \cref{c:r_dual} is established.

\end{itemize}

To show (1) implies (2), define $\imp^\star$ as in the Theorem: i.e., by \eqref{eq:impstar-def-inner} and \eqref{eq:impstar-def-outer} (across languages) and by $\imp_1$ and $\imp_2$ within languages.
That $\imp^\star$ is well defined is precisely the content of \cref{c:galois}.
We will now show the required properties of $\imp^\star$:

\begin{itemize}
  \item[\iref{i:extensibility}:] Immediate from the definition.
  \item[\iref{i:transitivity}:] Let $\lambda \imp^\star \eta \imp^\star \iota$. We must show that $\lambda \imp^\star \iota$. There are various cases depending on from which language these propositions arise. If $\lambda,\eta,\iota \in \LL_i$, the result follows from the transitivity of $\imp_i$. 
  For each case, the subscript on the proposition indicates which language it comes from, assuming $i \neq j$:
  \begin{itemize}
    \item Case: $\lambda_i \imp^\star \eta_j \imp^\star \iota_j$. Then $\Tij^{+}(\lambda_i) \imp_j \eta_j$ so by transitivity of $\imp_j$ we have $\Tij^{+}(\lambda_i) \imp_j \iota_j$. By definition $\lambda_i \imp^\star \iota_j$.
    \item Case: $\lambda_i \imp^\star \eta_i \imp^\star \iota_j$. Then $\eta_i \imp_i \Tji^{-}(\iota_j)$ so by transitivity of $\imp_j$ we have $\lambda_i \imp_i \Tji^{-}(\iota_j)$. By definition $\lambda_i \imp^\star \iota_j$.
    \item Case: $\lambda_j \imp^\star \eta_i \imp^\star \iota_j$. Then $\lambda_j \imp_j \Tij^{-}(\eta_i)$ and $\Tij^{+}(\eta_i) \imp_j \iota_j$. By \cref{c:approximation}, $\Tij^{-}(\eta_i) \imp_j \Tij^{+}(\eta_i)$, so by the transitivity of $\imp_j$, $\lambda_j \imp_j \iota_j$ and therefore also $\lambda_j \imp^\star \iota_j$.
  \end{itemize}
        \item[\iref{i:explosion}:] Immediately follows from \tref{t:contradiction} and Proposition \ref{prop:equivalence}. 

      \item[\iref{i:connective}:] We will show (i), where (ii) is analogous. 
       Let $\lambda_i \imp^{\star} \eta_j$ and $\lambda_i \imp^{\star} \eta'_j$. Then we have $\Tij^{+}(\lambda_i) \imp_j \eta_j$ and $\Tij^{+}(\lambda_i) \imp_j \eta'_j$; it follows that $\Tij^{+}(\lambda_i) \imp_j \eta_j \land \eta'_j$, and so by definition $\lambda_i \imp^\star \eta_j \land \eta'_j$

      \item[\iref{i:neg_cons}:] 
        Fix $\lambda_i \in \LL_i$ and $\eta_j \in \LL_j$ such that $\lambda_i \imp^\star t_j$ and $\eta_j \imp^\star \neg \lambda_i$. So from \eqref{eq:impstar-def-inner} we have $\lambda_i \imp_i \Tji^{-}(t_j)$ and $\eta_j \imp_j \Tij^{-}(\neg \lambda_i)$.
        So by \cref{c:r_dual}, we have $\eta_j \imp_j \neg \Tij^{+}(\lambda_i) \land \Tij^{-}(t_i)$; in particular this implies $\eta_j \imp_j \neg \Tij^{+}(\lambda_i)$, or since $\imp_j$ is Boolean, $\Tij^{+}(\lambda_i) \imp_j \neg \eta_j$. \eqref{eq:impstar-def-outer} delivers $\lambda_i \imp^\star \neg \eta_j$.
 \end{itemize}

Uniqueness is immediate.

Now assume in addition that $\L_j$ is complete. Let $H_j = \{ \eta_j \in \LL_j \mid \eta_j \imp^\star \lambda_i \}$. By Lemma \ref{lem:selfimply}, we have $\bigvee H_j \imp^\star \lambda_i$, and so, by \eqref{eq:impstar-def-inner}, $\bigvee H_j \imp_j \Tij^{-}(\lambda_i)$.

Let 
$$\eta^\dag_j = \Tij^{-}(\lambda_i) \land \neg \bigvee H_j.$$
Thus, $\eta^\dag_j \imp_j  \Tij^{-}(\lambda_i)$ and so, applying \eqref{eq:impstar-def-inner}, we obtain $\eta^\dag \imp^\star \lambda_i$, indicating that $\eta^\dag_j \in H_j$. 

Further $\eta^\dag_j \imp_j  \neg \bigvee H_j$, or equivalently, $\bigvee H_j \imp_j \neg \eta^\dag_j$: by definition of supremum this means that for all $\eta_j \in H_j$, $\eta_j \imp_j \neg \eta^\dag_j$, and so (since $\eta^\dag_j \in H_j$) we have that $\eta^\dag_j \imp_j \neg \eta^\dag_j$, or that $\eta^\dag_j = f_j$. So  $\Tij^{-}(\lambda_i) \imp_j \bigvee H_j$. 

The case for $\Tij^{+}$ is equivalent.
\end{tproof}

\begin{pproof}{prop:semantic-to-trans}
(1) implies (2): We will show this is the case an inner operator $\Tij^{-}$, where the case for the outer operator is the same:
\begin{align*}
\v_j \circ \Tij^{-} \circ \v_i^{-1} (E) &=  \v_j\big(\bigvee \{ \eta \in \LL_j \mid \eta \imp^\star \v_i^{-1}(E) \}\big) \\
&=  \v_j\big(\bigvee \{ \eta \in \LL_j \mid \v_j(\eta) \subseteq E \}\big) \\
&=  \bigcup \{ \v_j(\eta) \mid \eta \in \LL_j, \v_j(\eta) \subseteq E \} \\
&=  \bigcup \{ F \in \Sigma_{j} \mid F \subseteq E \}
\end{align*}

(2) implies (3): this follows immediately from (pre)-composition with $\v_j^{-1}$ and (post)-composition with $\v_i$.

(3) implies (1): Let $\eta \imp \lambda$. The only non-trivial case is where $\lambda \in \LL_i$ and $\eta \in \LL_j$ with $i\neq j$.
We have $\eta \imp^\star \lambda$ if and only if  $\eta \imp^\star \Tij^{-}(\lambda)$ (by \eqref{eq:impstar-def-inner}) if and only if $\v_j(\eta) \subseteq \v_j(\Tij^{-}(\lambda))$. Moreover we have $\v_j(\Tij^{-}(\lambda)) = \Vji(\v_i(\lambda)) = \bigcup\{F \in \Sigma_j \mid F \subseteq \v_i(\lambda)\}$. Clearly,  $\v_j(\eta) \subseteq \bigcup\{F \in \Sigma_j \mid F \subseteq \v_i(\lambda)\}$ if and only if $\v_j(\eta) \subseteq \v_i(\lambda)$, completing the proof.
\end{pproof}


\begin{tproof}{thm:state-space-existance}
\textbf{Preliminaries.} If $X$ is a Boolean algebra, with $\imp$ the canonical ordering call $F \subseteq X$ a \emph{$X$-filter}, (or \emph{filter} when the algebra is obvious from the context) if:
\begin{itemize}[topsep=0pt,itemsep=-1ex]
  \item $F$ is non-empty
  \item $F$ is $\imp$-upwards-closed
  \item  $x,y \in F$ then $x \land y \in F$. 
\end{itemize}
A filter is called \emph{proper} if it is not the totality of $X$.  
Let $\mathfrak{F}(X)$ collect all proper filters over $X$. It is easy to verify that $\mathfrak{F}(X)$ is closed under intersections. A proper filter is called an \emph{ultra}filter if it cannot be extended (by set inclusion) to any other proper filter. Let $\mathfrak{U}(X)$ collect all ultrafilters over $\imp$.  
It is well know that proper filter $F$ is an ultrafilter if and only if it is \emph{prime}: $a \lor b \in F$ implies $a \in F$ or $b \in F$. In particular, proper filter $F$ is an ultrafilter if and only if for all $a \in F$ either $a \in F$ or $\neg a \in F$.

For a subset $Z \subseteq \LS_1 \cup \LS_2$, let $Z_i = Z \cap \LL_i$ for $i \in \{1,2\}$. For $Z \subseteq \LS_1 \cup \LS_2$ let $\up(Z) = \{\eta \in \LS_1 \cup \LS_2 \mid \lambda \imp^\star \eta, \lambda \in Z\}$ denote the set of all propositions (in either language and including $\ast$) that are $\imp^\star$ implied by the propositions in $Z$.

Call a subset $\omega \subset \LS_1 \cup \LS_2$ a \emph{pre-state} if 
\begin{itemize}[topsep=0pt,itemsep=-1ex]
  \item $\w$ is non-empty
  \item $\w$ is $\imp^\star$-upwards-closed
  \item for both $i \in \{1,2\}$ we have $\omega_i$ is either empty or a proper $\LL_i$ filter.
\end{itemize}
and a \emph{state} if the last condition is replaced by
\begin{itemize}[topsep=0pt,itemsep=-1ex]
  \item for both $i \in \{1,2\}$ we have $\omega \cap L_i$ is either empty or an $\LL_i$ ultrafilter.
\end{itemize}

\begin{lemma}
\label{lem:prestate_extension_upset}
Let $\imp^\star$ satisfy \iref{i:extensibility}--\iref{i:neg_cons}. If $F \in \mathfrak{F}(\LL_i)$, then $\up(F)$ is a pre-state.
\end{lemma}

\begin{subproof}
That $\up(F)$ is non-empty and $\imp^\star$-upwards closed are obvious. By \iref{i:extensibility}, we have that $(\up(F))_i = F$. So we need only show that $(\up(F))_j$ is a proper filter. Let $\eta_j, \eta'_j \in (\up(F))_j$; it follows that there exists some $\lambda_i,\lambda'_i \in F$ such that $\lambda_i \imp^\star \eta_j$ and $\lambda'_i \imp^\star \eta'_j$. Since $F$ is a filter $\lambda_i \land \lambda'_i \in F$. By \iref{i:connective}, $\lambda_i \land \lambda'_i \imp^\star \eta_j \land \eta'_j$, so $\eta_j \land \eta'_j \in (\up(F))_j$. Moreover, if $f_j \in (\up(F))_j$, then there exists some $\lambda_i \in F$ such that $\lambda_i \imp^\star f_j$; by \iref{i:explosion} and \iref{i:extensibility}, this implies $\lambda_i \imp_i f_i$, contradicting the properness of $F$.
\end{subproof}

\begin{lemma}
\label{lem:prestate_extension_to_ultra}
Let $\imp^\star$ satisfy \iref{i:extensibility}--\iref{i:neg_cons}. Let $i,j \in \{1,2\}$ with $i \neq j$ and $F \in \mathfrak{F}(\LL_i)$ and $\eta_j \in \LL_j$ such that $\eta_j \notin \up(F)$. Then there exists an ultrafilter $U \in \mathfrak{U}(\LL_i)$ such that $\eta_j \notin \up(U)$.
\end{lemma}

\begin{subproof}
Order $\{F' \in \mathfrak{F}(\LL_i) \mid F \subseteq F', \eta_j \notin \up(F')\}$ by set inclusion. By Zorn's Lemma, there exists a maximal element of this partial order, call it $F^\dag$. We will show that $F^\dag$ an ultrafilter. Assume by way of contradiction that  $\lambda_i \notin F^\dag$ and $\neg \lambda_i \notin F^\dag$.

Consider 
\begin{align*}
  F^{\lambda} &= \{\phi_i \land \psi_i \mid \phi_i \in \LL_i \text{ such that } \lambda_i \imp_i \phi_i \text{ and } \psi_i \in F^\dag \}  \subseteq \LL_i, \text{ and } \\
    F^{\neg \lambda} &= \{\phi'_i \land \psi'_i \mid \phi'_i \in \LL_i \text{ such that } \neg \lambda_i \imp_i \phi'_i \text{ and } \psi'_i \in F^\dag \}  \subseteq \LL_i.
\end{align*}
That $F^{\lambda}$ and $F^{\neg \lambda}$ are a $\LL_i$-filters is immediate. They are also proper, as we will show for $F^{\lambda}$: Indeed, let $\psi_i \in F^\dag$. Since $\neg \lambda_i \notin F^\dag$, we have that $\psi_i \ {\centernot{\imp}}_i \ \neg \lambda_i$, or equivalently via the rules of Boolean algebra, $\lambda_i \ {\centernot{\imp}}_i\ \neg \psi_i$. Thus for any $\phi_i \in \LL_i$ such that $\lambda_i \imp_i \phi_i$ we have also $\phi_i \ {\centernot{\imp}}_i\ \neg \psi_i$, and hence $\phi_i \land \psi_i \neq f_i$.

So, since $F^{\lambda}$ and $F^{\neg \lambda}$ are proper filters containing $F^\dag$, by the assumed maximality of $F^\dag$, it must be that $\eta_j \in \up(F^{\lambda})$ and $\eta_j \in \up(F^{\neg \lambda})$. So, there must exist $\phi_i, \phi'_i, \psi_i,\psi'_i \in \LL_i$ with $\lambda_i \imp_i \phi_i$, $\neg\lambda_i \imp_i \phi'_i$ and $\psi_i, \psi'_i \in F^\dag$ such that
$$
(\phi_i \land \psi_i) \imp^\star t_j \qquad \text{ and } \qquad (\phi'_i \land \psi'_i) \imp^\star t_j
$$
and so via \iref{i:extensibility}
$$
(\phi_i \land \psi_i) \lor (\phi'_i \land \psi'_i) \imp^\star \eta_j
$$
Now notice that since both $\lambda \imp_i (\phi_i \lor \phi'_i)$ and $\neg \lambda \imp_i (\phi_i \lor \phi'_i)$, it follow that $\lambda_i \lor \neg \lambda_i \imp_i (\phi_i \lor \phi'_i) = t_i$. It follows that
\begin{align*}
  (\phi_i \land \psi_i) \lor (\phi'_i \land \psi'_i)  &= ((\phi_i \land \psi_i) \lor \phi'_i) \land ((\phi_i \land \psi_i) \lor \psi'_i) \\
  &= ((\phi_i  \lor \phi'_i) \land (\psi_i \lor \phi'_i)) \land ((\phi_i \lor \psi'_i)\land (\psi_i \lor \psi'_i)) \in F^\dag
\end{align*}
contradicting the assumption that $\eta_j \notin \up(F^\dag)$.
\end{subproof}

\begin{lemma}
\label{lem:prestate_extension_final}
Let $\imp^\star$ satisfy \iref{i:extensibility}--\iref{i:neg_cons}. Let $i,j \in \{1,2\}$ with $i \neq j$ and $\w$ be a pre-state such that $\w_i \in \mathfrak{F}(\LL_i)$. Moreover, assume $t_j \in \up(w_i)$ and $\eta_j \in \LL_j$ with $\neg \eta_j \notin \w$. Then there exists a pre-state $\w'$ such that $\w \subseteq \w'$ and $\eta_j \in \w'$.
\end{lemma}

\begin{subproof}
Let $\w$ and $\eta_j \in \LL_j$ be as given. 
Set
\begin{align*}
  P_i = \{\phi_i \land \psi_i \mid \phi_i \in \LL_i \text{ such that } \eta_j \imp^\star \phi_i \text{ and } \psi_i \in \w_i \}  \subseteq \LL_i \\
  P_j = \{\phi_j \land \psi_j \mid \phi_j \in \LL_i \text{ such that } \eta_j \imp^\star \phi_j \text{ and } \psi_j \in \w_j \}  \subseteq \LL_j 
\end{align*}
and consider the set of propositions:
$$
\w' = \w \cup P_i \cup P_j
$$

We will show that $\w'$ is a pre-state. That $\w'$ non-empty and $\imp^\star$-upwards closed are obvious; thus it suffices to show that $\w'_i$ and $\w'_j$ are proper filters.

We will show this for $\w'_i$; the argument for $\w'_j$ follows classical arguments in constructing ultrafilters.

If $\eta_j \ {\centernot{\imp}}^\star\  t_i$ then $P_i = \emptyset$ (i.e., there are no $\phi_i \in \LL_i$ such that  $\eta_j \imp^\star \phi_i$): it follows that $\w'_i = \w_i$ and we are done. So assume $\eta_j \imp^\star  t_i$, we have that $\phi_i = \phi_i \land t_i \in P_j$ for all $\phi_i \in \w_i$, and so, $\w_i \subseteq P_i$; we must show $P_i$ is a proper filter. 

That $P_i$ is non-empty is obvious. Let $\lambda_i \in \LL_i$ be such that $(\phi_i \land \psi_i) \imp_i \lambda_i$ for some $\phi_i, \psi_i \in \LL_i$ with $\eta_j \imp^\star \phi_i$ and $\psi_i \in \w_i$. Then $\eta_j \imp^\star (\phi_i \lor \lambda_i)$ and $(\psi_i\lor \lambda_i) \in \w_i$, so $\lambda_i = (\phi_i \land \psi_i) \lor \lambda_i = (\phi_i \lor \lambda_i) \land (\psi_i \lor \lambda_i) \in P_i$. So $P_i$ is $\imp_i$-upwards closed. 

Next, let $(\phi_i \land \psi_i), (\phi'_i \land \psi'_i) \in P_i$ so that $\eta_j \imp^\star \phi_i$ and $\eta_j \imp^\star \phi'_i$ and $\psi_i,\psi'_i \in \w_i$. Since $\w_i$ is a proper filter, $(\psi_i \land \psi'_i) \in \w_i$ and by \iref{i:connective}, $\eta_j \imp^\star (\phi_j \land \phi'_j)$, so $(\phi_i \land \phi'_i) \land (\psi_i \land \psi'_i) = (\phi_i \land \psi_i) \land (\phi'_i \land \psi'_i) \in P_i$. $P_i$ is directed, hence a filter.

We will now show properness.  Assume by way of contradiction that $f_i \in P_i$, so, there must exist $\phi_i, \phi'_i, \psi_i,\psi'_i \in \LL_i$ with $\eta_j \imp^\star \phi_i$ and $\eta_j \imp^\star \phi'_i$ and $\psi_i,\psi'_i \in \w_i$ such that 
\begin{align*}
f_i &= (\phi_i \land \psi_i) \land (\phi'_i \land \psi'_i)\\
&= (\phi_i \land \phi'_i) \land (\psi_i \land \psi'_i)
\end{align*}
We can re-write this as $(\psi_i \land \psi'_i) \imp_i \neg (\phi_i \land \phi'_i)$.
Since $\w_i$ is a proper filter, $(\psi_i \land \psi'_i) \in \w_i$; by \iref{i:connective} $\eta_j \imp^\star (\phi_j \land \phi'_j)$

By assumption, $t_j \in \up(w_i)$, so let $\lambda'_i \in w_i$ be such that $\lambda'_i \imp^\star t_j$. Let $\lambda_i = \lambda'_i \land (\psi_i \land \psi'_i)$; since $\w_i$ is an ultrafilter we have $\lambda_i \in \w_i$.
So, we have
$$
\eta_j \imp^\star (\psi_i \land \psi'_i) \imp^\star \neg (\phi_i \land \phi'_i) \imp^\star \neg \lambda_i
$$
Thus by \iref{i:neg_cons} we have $\lambda_i \imp^\star \neg \eta_j$, contradicting our assumption that $\neg \eta_j \notin \w$.
\end{subproof}

We will now prove the theorem:

(\ref{thm:sse:T} $\rightarrow$ \ref{thm:sse:i}) 
Follows from Theorem \ref{th:imp-equivalence}.

(\ref{thm:sse:i} $\rightarrow$ \ref{thm:sse:state-space}) Let $\W$ denote the set of \emph{states}, i.e., non-empty directed subsets of $\LS_1 \cup \LS_2$ such that $w \cap \LL_i$ is empty or an $\LL_i$ ultrafilter for $i \in \{1,2\}$.

Let $\v_i: \lambda_i \mapsto \{w \in \W \mid \lambda_i \in \W\}$ (and $\v_i(\ast) = \ast$) denote the canonical evaluation map. From the properties of ultrafilters, is obvious that $\v_i$ preserves conjunction and relative negation.  Thus, in light of Proposition \ref{prop:semantic-to-trans}, it suffices to show, where we allow the possibility that $i=j$, that $\lambda_i \imp^\star \eta_j$ if and only if $\v_i(\lambda_i) \subseteq \v_j(\eta_j)$. If $\lambda_i \imp^\star \eta_j$ it follows immediately from the fact that states are $\imp^\star$-upward closed that $\v_i(\lambda_i)\subseteq \v_j(\eta_j)$.

We will now show the `only-if' direction: assume that  $\lambda_i \ {\centernot{\imp}}^\star\ \eta_j$. By \iref{i:explosion}, we know $\lambda_i \neq f_i$. Let $F = \{\lambda'_i \in \LL_i \mid \lambda_i \imp_i  \lambda'_i\}$. It is immediate that $F \in \mathfrak{F}(\LL_i)$ and $\eta_j \notin \up(F)$.
We claim that there exists an ultrafilter, $U \in \mathfrak{U}(\LL_i)$, such that $F \subseteq U$ and $\eta_j \notin \up(U)$. If $i = j$, this follows form standard constructions of ultrafilters. If $i \neq j$, this is the content of Lemma \ref{lem:prestate_extension_to_ultra}.

Let $\w = \up(U)$; by Lemma \ref{lem:prestate_extension_upset}, $\w$ is a pre-state such that $w_i \in \mathfrak{U}(\LL_i)$. Let $k \in \{1,2\}$ with $k \neq i$. If $\w_k$ is empty then $\w$ is a state, and so $\w \in \v_j(\eta_j) \setminus \v_i(\lambda_i)$. So assume $\w_k$ is non-empty, implying $t_k \in \w$.  If $i=j$, then since $w_i$ is an ultrafilter, it follows that $\neg\eta_j \in w$. If $i \neq j$, then by Lemma \ref{lem:prestate_extension_final}, we an extend $\w$ to a pre-state $\w'$ such that $w\subseteq w'$ such that $\neg\eta_j \in w'$. 

Consider the set of pre-states $\{\w' \text{ is pre-state} \mid w' \subseteq \w, \neg \eta_j \in \w'\}$, which is non-empty by the above argument. Order this set of pre-states by set inclusion and appeal to Zorn to select an maximal element, $\w^\dag$. Clearly, $w^\dag_i = w_i$ is an ultrafilter. Furthermore, $w^\dag_k$ must be an ultra-filter. Towards a contradiction, suppose not: then there exists some $\lambda_k \in \LL_k$ such that $\lambda_k \notin w^\dag_k$ and $\neg\lambda_k \notin w^\dag_k$. Then we can use Lemma \ref{lem:prestate_extension_final} to obtain a proper extension of $w^\dag$, a contradiction to its maximality. So $\w^\dag$ is a state and moreover $\w^\dag  \in \v_j(\eta_j) \setminus \v_i(\lambda_i)$.

(\ref{thm:sse:state-space} $\rightarrow$ \ref{thm:sse:T}) Define $\V := \v_j \circ \T \circ \v_i^{-1}$.  For any $\eta_j$ and $\lambda_i$ we have
\begin{align*}
\eta_j \imp_j \Tij^{-}(\lambda_i) &\qquad \text{iff} \qquad \eta_j \imp_j \v_j^{-1} \circ \Vij^{-} \circ \v_i (\lambda_i) \\
&\qquad \text{iff} \qquad  \v_j(\eta_j) \subseteq \Vij^{-} \circ \v_i (\lambda_i)\\
&\qquad \text{iff} \qquad  \v_j(\eta_j) \subseteq  \bigcup \Big\{F \in \Sigma_{j} \mid F \subseteq \v_i (\lambda_i) \Big\} \\
&\qquad \text{iff} \qquad  \v_j(\eta_j) \subseteq  \v_i (\lambda_i) \\
&\qquad \text{iff} \qquad  \bigcap \Big\{F \in \Sigma_{i} \mid \v_j(\eta_j) \subseteq F\Big\}  \subseteq  \v_i (\lambda_i) \\
&\qquad \text{iff} \qquad  \v_i^{-1} \circ\Vji^{+} \circ \v_j(\eta_j) \imp_i  \lambda_i\\
&\qquad \text{iff} \qquad  \Tji^{+}(\eta_j) \imp_i  \lambda_i
\end{align*}
Hence \cref{c:galois} holds. The other two properties are equally straightforward applications of the definition of $\V$.
\end{tproof}

\begin{pproof}{lemma-common-lang-exist} Fix $\lambda \in \LL_i$, then

\begin{itemize}
  \item[($1 \implies 2$)] Assume $\Tij^{-}(\lambda) = \Tij^{+}(\lambda)$. Call this element $\eta_\lambda$. By reflexivity of $\imp_j$ we have: $\Tij^{+}(\lambda) \imp_j \eta_\lambda \imp_j \Tij^{-}(\lambda)$. Then applying \eqref{eq:impstar-def-outer} (to the first implication) and \eqref{eq:impstar-def-inner} (to the second), we obtain $\lambda \imp^{\star} \eta_\lambda \imp^{\star} \lambda$, as desired. 

  \item[($2 \implies 3$)] Assume $\lambda \imp^{\star} \eta_\lambda \imp^{\star} \lambda$. Then by \eqref{eq:impstar-def-outer} (to the first implication) $\Tij^{+}(\lambda) \imp_j \eta_\lambda$, and so by \iref{i:extensibility} and \iref{i:transitivity}, $\Tij^{+}(\lambda)  \imp^{\star} \lambda$ \eqref{eq:impstar-def-outer} then yields $\Tji^{+}(\Tij^{+}(\lambda))  \imp_i \lambda$. That $\lambda \imp_i \Tji^{+}(\Tij^{+}(\lambda))$ follows from Lemma \ref{lem:gal-implications}\ref{lem:imps:closure} and \tref{t:monotonicity}.

  \item[($3 \implies 4$)] Assume $\Tji^{+}(\Tij^{+}(\lambda)) = \lambda$. Then $\Tji^{+}(\Tij^{+}(\lambda))  \imp_i \lambda$; applying \cref{c:galois} twice yields $\lambda \imp_i \Tji^{-}(\Tij^{-}(\lambda))$. $\Tji^{-}(\Tij^{-}(\lambda)) \imp_i \lambda $ follows from Lemma \ref{lem:gal-implications}\ref{lem:imps:interior} and \tref{t:monotonicity}.

  \item[($4 \implies 1$)] Assume $\Tji^{-}(\Tij^{-}(\lambda)) = \lambda$. Then $\lambda \imp_i \Tji^{-}(\Tij^{-}(\lambda))$; applying \eqref{eq:impstar-def-inner} provides $\lambda \imp^{\star} \Tij^{-}(\lambda)$; applying \eqref{eq:impstar-def-inner} again provides $\Tij^{+}(\lambda) \imp_j \Tij^{-}(\lambda)$. That $\Tij^{-}(\lambda) \imp_j \Tij^{+}(\lambda)$ follows from \cref{c:approximation}.

\end{itemize}
 

 So let $\L = \{\lambda \in \LL_i \mid \lambda \Leftrightarrow^\ast \eta_\lambda, \text{ for some } \eta_\lambda \in \LL_j\}$. We will show that $\L$ is closed under conjunction and relative complementation, and contains a maximal and minimal element (which are distinct since there are at least two elements). It is thus a Boolean algebra under the inherited operations. This completes the proof as taking $\epsilon_i$ to be the identity and $\epsilon_j: \lambda \mapsto \eta_\lambda$ serves to show that $\L$ is a common language. 

That $f_i \in \L$ is an immediate consequence of \iref{i:explosion}. Let $\bar{\lambda} = \bigvee \{\lambda \in \LL_i \mid \lambda \in \L\}$ and $\bar{\eta} = \bigvee \{\eta_\lambda \in \LL_j \mid \lambda \in \L\}$. Applying \iref{i:connective} in both directions, we have $\bar{\lambda} \Leftrightarrow^\ast \bar{\eta}$, so $\bar{\lambda} \in \L$ is a maximal element. 

Now, let $\lambda, \lambda' \in \L$. Applying \iref{i:connective} twice shows that $(\lambda \land \lambda') \Leftrightarrow^\ast (\eta_{\lambda} \land \eta_{\lambda'})$; so $\L$ is closed under conjunctions. 

We have $(\neg \lambda \land \lambda') \imp^{\star} \lambda' \imp^{\star} \eta_{\lambda'} \imp^{\star} t_j$, and $\eta_\lambda \imp^{\star} \lambda \imp^{\star} (\lambda \lor \neg \lambda') = \neg (\neg \lambda \land \lambda')$. So \iref{i:neg_cons} yields $(\neg \lambda \land \lambda') \imp^{\star} \neg \eta_\lambda$. Thus by \iref{i:connective}, $(\neg \lambda \land \lambda') \imp^{\star} (\neg \eta_\lambda \land \eta_{\lambda'})$. A symmetric argument furnishes the opposite implication; $\L$ is closed under relative complementation.
\end{pproof}

\newpage
\singlespacing

\setlength{\bibsep}{1pt plus 0.3ex}

\bibliographystyle{aer}
\bibliography{it.bib}

\end{document}